\documentclass[aps,showpacs,twocolumn,superscriptaddress,groupedaddress]{revtex4} 
   
\usepackage{amssymb}
\usepackage{amsmath}
\usepackage{graphicx}
\usepackage{subfigure}
\usepackage{color}
\usepackage{mathrsfs}
\usepackage[dvipsnames]{xcolor}
\usepackage[breaklinks=true,colorlinks=true]{hyperref}
\hypersetup{colorlinks=true,citecolor=blue,linkcolor=blue,urlcolor=blue}
\usepackage[utf8]{inputenc}
\usepackage[english]{babel}

\begin{document}
\title{Nonlinear Dynamics of Kink Configurations: \\From Small to Large Kink Collisions}

\author{Aliakbar Moradi Marjaneh}
\email{moradimarjaneh@iau.ac.ir, moradimarjaneh@gmail.com}
\affiliation{Department of Physics, Qu.C., Islamic Azad University, Quchan, Iran.}

\author{Dionisio Bazeia}
\email{bazeia@fisica.ufpb.br, dbazeia@gmail.com}
\affiliation{Department of Physics, Federal University of Paraíba, João Pessoa, Paraíba, Brazil.}

\begin{abstract}
This study explores the scattering dynamics of kinks within a nonlinear system governed by a parameterized potential $U_\lambda(\chi)$, examining the distinct behaviors of small and large kinks across a range of $\lambda$ values and initial velocities. For small kinks, we investigate the critical velocity for separation, the influence of vibrational modes, resonance phenomena, and the conditions under which large kinks emerge from collisions. Our findings reveal that the critical velocity exhibits a non-monotonic dependence on the parameter $\lambda$, reflecting the evolving stability of small kinks, while the decreasing frequency of vibrational modes with increasing $\lambda$ diminishes resonance effects, leading to simpler scattering dynamics at higher $\lambda$. The formation of large kinks from small kink collisions is favored at lower $\lambda$, where the mass difference between small and large kinks is reduced. Conversely, large kink scattering consistently results in the production of small kinks, with the number of small kink pairs growing as both $\lambda$ and initial velocity increase, a process driven by energy transfer from the translational modes of large kinks to the potential energy required for small kink creation. The absence of vibrational modes in large kinks contrasts with their presence in small kinks, where such modes give rise to complex phenomena like bion formation and resonance. These results underscore the pivotal role of $\lambda$ in shaping kink interactions and offer valuable insights into the dynamics of topological defects in nonlinear systems, with potential implications for understanding similar phenomena in condensed matter physics and related fields.
\end{abstract}



\maketitle

\section{Introduction}\label{sec:introduction}
Topological solutions such as kinks and antikinks in (1+1)-dimensional scalar field theories are powerful tools for studying nonlinear phenomena in physics due to their stability and unique structure \cite{Rajaraman.book.1982,Vilenkin.book.2000,Manton.book.2004,Shnir.book.2018}. These configurations find applications across diverse fields, ranging from early universe cosmology \cite{Vilenkin.book.2000} to condensed matter physics and nonlinear optics \cite{Bishop.PhysD.1980,Optics2}. In integrable models such as the sine-Gordon model, kink-antikink scattering exhibits simple elastic behavior accompanied by a phase shift \cite{Dmitriev.Nonlinearity.2000,Dmitriev.PRE.2008,Moradi.EPJB.2018}. However, in nonintegrable models—such as the $\phi^4$\cite{Rajaraman.book.1982,Makhankov.PhysRep.1978,Campbell.PhysD.1983,Moradi.CNSNS.2017,Askari.CSF.2020,Manton.PRL.2021,Almeida.EPJC.2025}, $\phi^6$\cite{Moradi.JHEP.2017,Gani.JOP.2020,Saadatmand.EPJB.2022,Adam.PRD.2022,Saadatmand.CSF.2024}, higher order models \cite{Khare.PRE.2014, Blinov.JOP.2020,BCM, khare2021explicit}, deformed models such as \cite{BBG,Moradi.CSF.2022, Moradi.AnlPhys.2024}, and double sine-Gordon (DSG)~\cite{Campbell.dsG.1986,Peyravi.EPJB.2009,Gani.EPJC.2018,Zolotaryuk.EPJC.2018,Belendryasova.JPCS.2019,Yerin.PRB.2021} models—a richer dynamics emerges, including the production of new kink-antikink pairs, solitary oscillations, and resonance windows. These complex behaviors have made the study of scattering a key approach to understanding nonlinear effects in physical systems.

The versatility of nonlinear models with tunable parameters further enriches this landscape. For example, studies on one-dimensional chains with adjustable anharmonicity have shown that the interaction of phonons with discrete breathers exhibits a strong dependence on the type and strength of nonlinearity \cite{Hadipour.PLA.2020}. Such findings highlight the broader relevance of parameter-driven dynamics in nonlinear systems, extending beyond field theories to lattice models; see, e.g., Ref. \cite{DB-2022} and references therein for investigations concerning bifurcation and chaos in chains of small particles. Similarly, an intriguing aspect of field-theoretic investigations is the emergence of comparable outcomes across models with seemingly distinct potentials. The double sine-Gordon (DSG) model, explored in \cite{Simas.JHEP.2020}, demonstrates that kink-antikink scattering can lead to the production of multiple kink-antikink pairs and transient oscillations, a phenomenon tied to the topological structure and model parameters. Far from being a drawback, such similarities may indicate universal behaviors in nonlinear systems that transcend the specifics of individual potentials. This observation underscores the importance of examining new models to validate and extend these patterns, enhancing our understanding of their underlying mechanisms.

In another work introduced in Ref. \cite{Bazeia.EPJC.2017}, the investigation proposed a model derived from the deformation of the $\phi^4$ model, controlled by a parameter $\lambda$. This model enables a smooth transition between the sine-Gordon regime ($\lambda=0$) and vacuumless systems ($\lambda=1$), supporting two types of kinks (large and small). While the work focused on the stability of solutions and braneworld applications, the scattering of kinks and antikinks in this model remained unexplored. This gap serves as the main motivation for the present study. Here, we investigate the collision of kinks and antikinks in the model for the first time, with the aim of analyzing its topological dynamics as a function of $\lambda$ and initial velocity.

Our findings reveal similarities with behaviors reported in the DSG model, notably the production of small kink-antikink pairs during large kink scattering. This overlap not only validates the proposed model but also suggests that common mechanisms may govern such phenomena across a broader class of nonlinear systems, echoing the parameter-dependent dynamics observed in lattice models~\cite{Askari.CSF.2020,Hadipour.PLA.2020,Speight.Nonlinearity.1997,kevrekidis.PhysicaD.2003,Flach.PhysRep.2008,Zhanna.IOPconf.2018,dmitriev.JPA.2005}. Moreover, the study of these effects within the framework of the model described in Ref. \cite{Bazeia.EPJC.2017}, which uniquely bridges distinct topological regimes, may offer a fresh perspective to deepen our understanding of the role of model parameters in shaping the scattering dynamics.

This paper is organized as follows. In Section II, we introduce the model and its key features. In Section III, we describe the numerical methods to be used in Sections IV and V to present the results of small and large kink scattering, respectively. Finally, in Section VI, we summarize our findings and suggest directions for future research.

\section{The Model}\label{sec:model}
We explore a scalar field theory in (1+1)-dimensional spacetime, characterized by a real scalar field $\phi(x,t)$ interacting through a non-negative potential $V(\phi)$. This potential exhibits a set of minima, denoted as $\mathcal{V} = \{ \phi_1^{(vac)}, \phi_2^{(vac)}, \phi_3^{(vac)}, \dots \},$ that define the vacuum states of the system. The dynamics of the field are governed by the Lagrangian density \cite{Rajaraman.book.1982,Vilenkin.book.2000,Manton.book.2004,Shnir.book.2018}
\begin{eqnarray}\label{eq:lagrangian}
\mathcal{L}(\phi, \partial_\mu \phi) &=& \frac{1}{2} \partial_\mu \phi \partial^\mu \phi - V(\phi) \nonumber \\
&=& \frac{1}{2} \left( \frac{\partial \phi}{\partial t} \right)^2 - \frac{1}{2} \left( \frac{\partial \phi}{\partial x} \right)^2 - V(\phi),
\end{eqnarray}
from which the energy functional is derived as
\begin{eqnarray}\label{eq:energyfunctional}
E[\phi] = \int_{-\infty}^{+\infty} \left[ \frac{1}{2} \left( \frac{\partial \phi}{\partial t} \right)^2 + \frac{1}{2} \left( \frac{\partial \phi}{\partial x} \right)^2 + V(\phi) \right] dx.
\end{eqnarray}
The resulting equation of motion is \cite{Rajaraman.book.1982,Vilenkin.book.2000,Manton.book.2004,Shnir.book.2018}
\begin{equation}\label{eq:EOM}
\frac{\partial^2 \phi}{\partial t^2} - \frac{\partial^2 \phi}{\partial x^2} + \frac{dV}{d\phi} = 0,
\end{equation}
and for static configurations, $\phi = \phi(x)$, it reduces to
\begin{equation}\label{eq:EOMstatic}
\frac{d^2 \phi}{dx^2} = \frac{dV}{d\phi}.
\end{equation}
This second-order equation can be reformulated in terms of the first-order differential equations,
\begin{equation}\label{eq:EOMstaticfirstorder}
\frac{d\phi}{dx} = \pm \sqrt{2 V(\phi)},
\end{equation}
which facilitate the derivation of kink and antikink solutions. These topological structures interpolate between adjacent vacua, satisfying
\begin{equation}\label{eq:vacua1}
\phi(-\infty) = \lim_{x \to -\infty} \phi(x) = \phi_i^{(vac)}, \quad \phi_i^{(vac)} \in \mathcal{V},
\end{equation}
and
\begin{equation}\label{eq:vacua2}
\phi(+\infty) = \lim_{x \to +\infty} \phi(x) = \phi_j^{(vac)}, \quad \phi_j^{(vac)} \in \mathcal{V},
\end{equation}
ensuring finite energy for static configurations. Moving kinks with velocity $v$ are obtained via a Lorentz boost $x\to\gamma(x-vt)$ with the Lorentz factor $\gamma=1/\sqrt{1-v^2}$ \cite{Rajaraman.book.1982,Vilenkin.book.2000,Manton.book.2004,Shnir.book.2018}. In this case, the field configuration changes as follows
\begin{equation}\label{eq:movingkinks}
\phi(x,t) \to \phi\left( \frac{x - v t}{\sqrt{1 - v^2}} \right).
\end{equation}
To compute the mass of static kinks or antikinks, one substitutes these solutions into the energy functional (\ref{eq:energyfunctional}). 
Moreover, since the potential $V$ is a non-negative function of the field $\phi$, we can introduce an auxiliary function $W=W(\phi)$ such that 
\begin{equation}
V(\phi)=\frac12 \,W_\phi^2\, ,   
\end{equation}
where $W_\phi=dW/d\phi$. This changes the above first-order equations into 
\begin{equation}
\frac{d\phi}{dx}=\pm W_\phi\, .  
\end{equation}

Next, we introduce a transformation to explore a deformed model. This procedure was first presented in \cite{BLM}, and later used in other contexts in \cite{DA,DB,DC}, and here we briefly review the main steps. To do this, let us consider another field $\chi=\chi(x,t)$ that is controlled by the new Lagrangian
\begin{equation}\label{defmod}
\mathcal{L}(\chi, \partial_\mu \chi) = \frac{1}{2} \partial_\mu \chi \partial^\mu \chi - U(\chi),
\end{equation}
where $U(\chi)$ is another potential. However, if one introduces a deformation function $f=f(\chi)$, and use it to change the field $\phi$ to $\chi$ as $\phi\to f(\chi)$, we can define the potential $U(\chi$) in the form
\begin{equation}\label{newpot}
U(\chi) = \frac{V(\phi \to f(\chi))}{f_\chi^2},
\end{equation}
where $f_\chi=df/d\chi$. This allows us to use $W(\phi)$ to introduce another function, $\bar{W}(\chi)$, such that 
\begin{equation}
\bar{W}_\chi = \frac{W_\phi(\phi \to f(\chi))}{f_\chi},
\end{equation}
and write first-order equations for the new or deformed model
in the form
\begin{equation}\label{eq1}
\frac{d\chi}{dx} = \pm \bar{W}_\chi.
\end{equation}

We can  show that the total energy of the static solutions 
$\chi(x)$ is given by
\begin{equation}\label{Ebpsdeformed}
\bar{E}[\chi] = |\bar{W}(\chi(\infty)) - \bar{W}(\chi(-\infty))|.
\end{equation}
It is also possible to show that a static solution $\phi(x)$ of the starting model can be linked to a static solution $\chi(x)$ of the deformed model through the expression
\begin{equation}\label{sol}
\chi(x) = f^{-1}(\phi(x)),
\end{equation}
via the inverse of the deformation function. In this sense, the deformation procedure allows us to deform a given model, which supports kinklike configuration, to obtain another model, which also supports kinklike configuration, constructed from solution of the starting model with the use of some appropriate deformation function.

\begin{figure*}[t!]
\centering
   \subfigure[]{\includegraphics[width=0.45\textwidth,height=0.20\textheight]{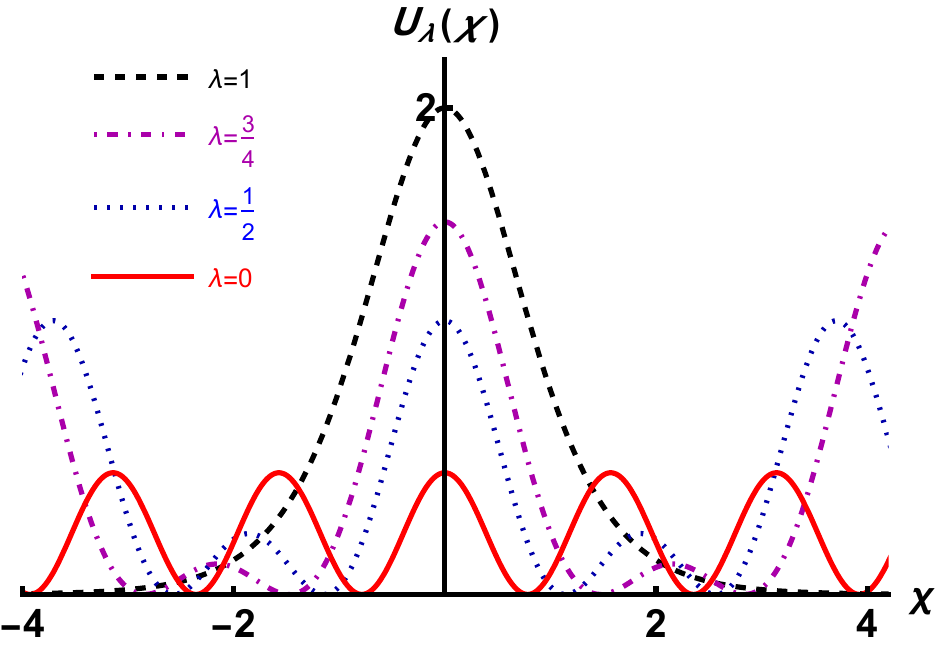}\label{fig:deformedpotential}} 
   \hspace{1mm} 
   \subfigure[]{\includegraphics[width=0.45\textwidth, height=0.20\textheight]{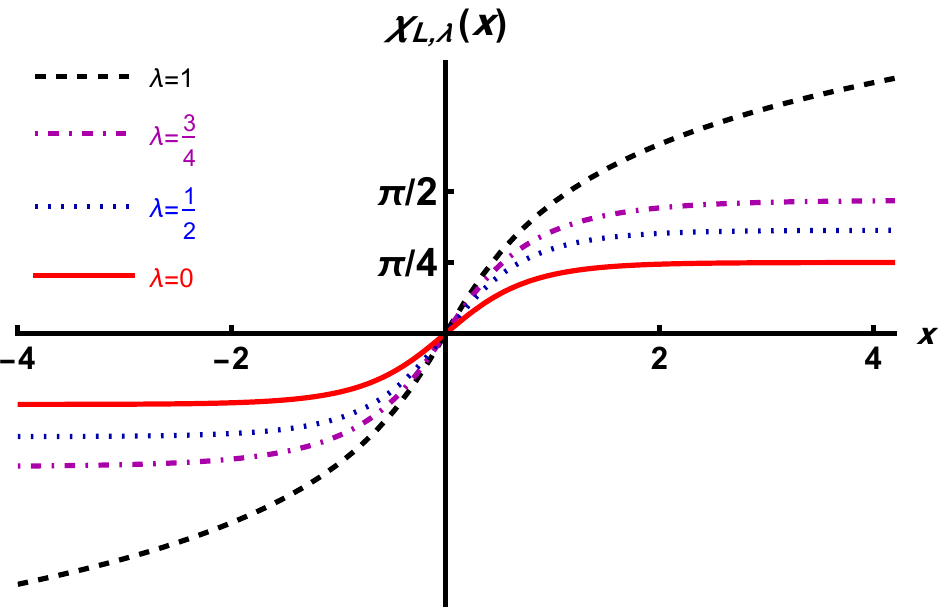}\label{fig:deformedlargekinks}}
   \hspace{1mm} \\
   \subfigure[]{\includegraphics[width=0.45\textwidth, height=0.20\textheight]{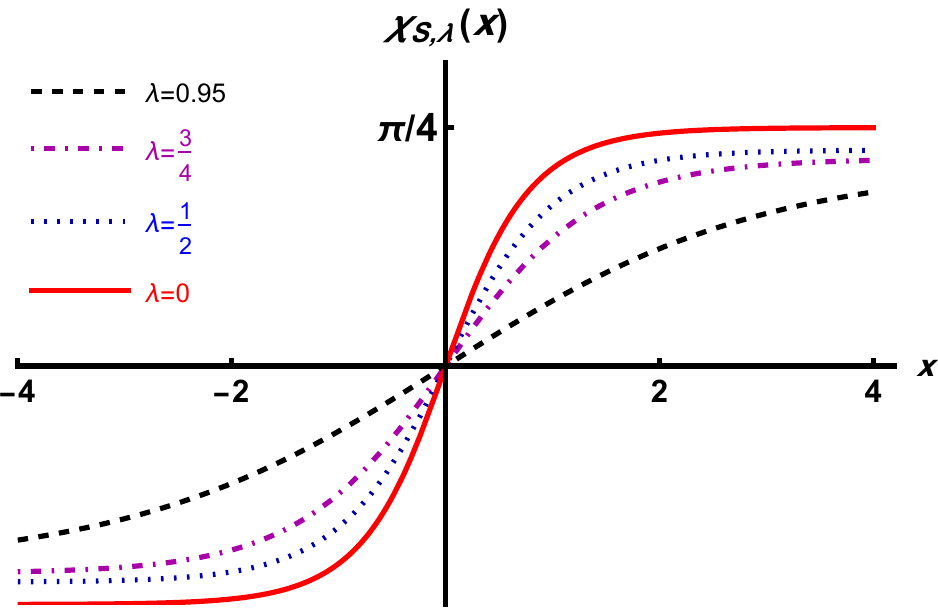}\label{fig:deformedsmallkinks}}
   \hspace{1mm} 
   \subfigure[]{\includegraphics[width=0.45\textwidth, height=0.20\textheight]{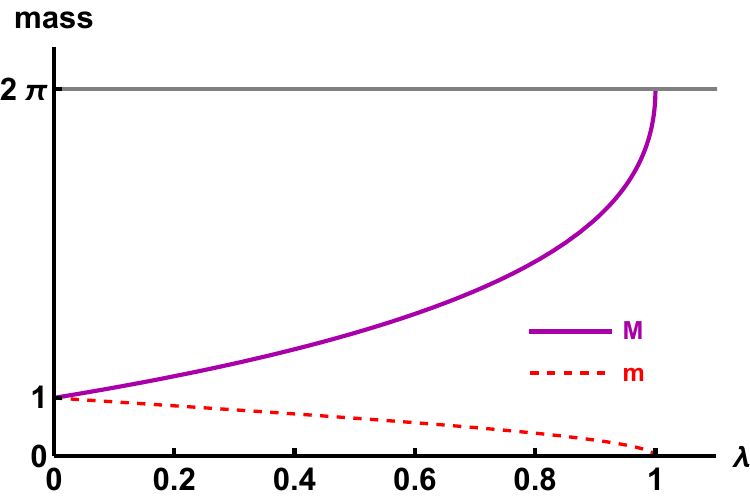}\label{fig:mass}}
\caption{(a) The deformed potential $U_\lambda(\chi)$ as defined in Eq.~\eqref{pot1} for different values of $\lambda$. (b) Spatial profiles of large kinks, $\chi_{L,\lambda}(x)$, from Eq.~\eqref{largek} for different values of $\lambda$. (c) Spatial profiles of small kinks, $\chi_{S,\lambda}(x)$, from Eq.~\eqref{smallk} for different values of $\lambda$. (d) Masses of small ($m$) and large ($M$) kinks as functions of $\lambda$, calculated using Eqs.~\eqref{smallkinkmass} and \eqref{largekinkmass}, respectively.}
\label{fig:kinkspotentialsmass}
\end{figure*}

Let us now start from the standard $\phi^4$ potential,
\begin{equation}\label{phi4}
V(\phi) = \frac{1}{2} (1 - \phi^2)^2,
\end{equation}
we apply the deformation function
\begin{equation}\label{d1}
f_\lambda(\chi) = \tanh\left( \frac{1}{\theta \sqrt{2 - \theta^2}} \text{tanh}^{-1} \left( \theta \frac{\text{sc}(\chi, \lambda)}{\sqrt{2 - \theta^2}} \right) \right),
\end{equation}
where $\theta$ is a real parameter ($\neq \pm \sqrt{2}$), and $\text{sc}(\chi, \lambda) = \frac{\text{sn}(\chi, \lambda)}{\text{cn}(\chi, \lambda)}$ involves Jacobi elliptic functions, with $\text{sn}(\chi, \lambda)$, $\text{cn}(\chi, \lambda)$, and $\text{dn}(\chi, \lambda)$ satisfying
\begin{subequations}\label{jacobirelations}
\begin{eqnarray}
\text{cn}^2(\chi, \lambda) + \text{sn}^2(\chi, \lambda) &=& 1, \\
\text{dn}^2(\chi, \lambda) + \lambda \text{sn}^2(\chi, \lambda) &=& 1.
\end{eqnarray}
\end{subequations}
The parameter $\lambda \in [0,1]$ tunes the model: at $\lambda = 0$, we recover trigonometric functions ($\text{sn}(\chi, 0) = \sin(\chi)$, $\text{cn}(\chi, 0) = \cos(\chi)$, $\text{dn}(\chi, 0) = 1$), while at $\lambda = 1$, hyperbolic functions emerge ($\text{sn}(\chi, 1) = \tanh(\chi)$, $\text{cn}(\chi, 1) = \text{dn}(\chi, 1) = \text{sech}(\chi)$). When required, we will use $\theta=\sqrt{1-\lambda\,}$. The resulting potential is shown in Fig.~\ref{fig:deformedpotential},
\begin{equation}\label{pot1}
 U_\lambda(\chi) = \frac{1}{2} \frac{\left( 2 \text{cn}^2(\chi, \lambda) - \theta^2 \right)^2}{\text{dn}^2(\chi, \lambda)}.
\end{equation}
Solutions to this model are found by solving
\begin{equation}\label{1eq}
\chi' = \frac{2 \text{cn}(\chi, \lambda)^2 - \theta^2}{\text{dn}(\chi, \lambda)}.
\end{equation}
This equation admits two distinct families of kink solutions. The first, termed \textit{large kinks}, is
{\small \begin{equation}\label{largek}
\chi_{L,\lambda}(x)\! = \text{sc}^{-1} \left( \frac{\sqrt{2 - \theta^2}}{\theta} \tanh\left( \theta \sqrt{2 - \theta^2} x \right), \lambda \right)\! + 2n \text{K}_\lambda.
\end{equation}}
\!Here $\text{K}_\lambda$ is the complete elliptic integral of the first kind. See Ref. \cite{Bazeia.EPJC.2017} for other details on this issue. 
For $n = 0$, the above solution, shown in Fig.~\ref{fig:deformedlargekinks}, connects asymptotic values $\chi_{L,\lambda}(\pm \infty) = \pm \text{sc}^{-1}(\sqrt{2 - \theta^2}/\theta, \lambda)$, with a near-origin behavior of $\chi_{L,\lambda}(x \simeq 0) \simeq (2 - \theta^2) x + \mathcal{O}(x^2)$. As $\theta \to 0$, the asymptotic limits diverge, while for $\theta \to 1$, the solution remains well-defined. At specific $\lambda$ values, it reduces to
\begin{subequations}
\begin{eqnarray}
\!\!\chi_{L,1}(x)\!\!\! &=&\!\!\! \text{sinh}^{-1}(2x) \quad\!\! \text{(vacuumless solution)},\; \\
\chi_{L,0}(x)\!\!\! &=&\!\!\! \text{tan}^{-1}(\tanh(x)) \quad\!\! \text{(sine-Gordon kink)}.\;
\end{eqnarray}
\end{subequations}
The second family, \textit{small kinks}, is given by
{\small \begin{equation}\label{smallk}
\chi_{S,\lambda}(x) = \text{sc}^{-1} \left( \frac{\theta \tanh \left( \theta \sqrt{2 - \theta^2} x \right)}{\sqrt{(2 - \theta^2)(1 - \lambda)}}, k \right) + (2n + 1) \text{K}_\lambda,
\end{equation}}
and is depicted in Fig.~\ref{fig:deformedsmallkinks}. 

The classical mass of a kink, defined as the energy of the static configuration, is
\begin{eqnarray}\label{mass}
\text{classical mass} = \int_{-\infty}^{+\infty} \left( \frac{d\chi}{dx} \right)^2 dx.
\end{eqnarray}
For small kinks,
\begin{eqnarray}\label{smallkinkmass}
m\!\! &=&\!\!\! \int_{-\infty}^{+\infty} \left( \frac{d\chi_{S,\lambda}(x)}{dx} \right)^2 dx \nonumber\\
&=&\!\! \frac{2}{\lambda}\!\! \left( 2 \tan^{-1}\!\! \left( \frac{\lambda - 1}{\sqrt{1 - \lambda^2}} \right)\! \!- R_\lambda \cot^{-1} \!\!\left( r_\lambda \right) \right)\!,
\end{eqnarray}
and for large kinks,
\begin{eqnarray}\label{largekinkmass}
M\!\! &=&\!\!\! \int_{-\infty}^{+\infty} \left( \frac{d\chi_{L,\lambda}(x)}{dx} \right)^2 dx\nonumber\\
&=& \frac{2}{\lambda} \left( 2 \tan^{-1} \left( \frac{\sqrt{\lambda + 1}}{\sqrt{1 - \lambda}} \right) + {R_\lambda} \tan^{-1} \left(r_\lambda\right) \right)\!,
\end{eqnarray}
where $R_\lambda=(\lambda^2+\lambda-2)/\sqrt{1-\lambda}$ and $r_\lambda=\sqrt{\lambda+1}$. 
These masses, plotted in Fig.~\ref{fig:mass}, show that $M > m$ for all $\lambda \in (0,1)$. This disparity suggests that large kink collisions possess sufficient energy to produce small kink-antikink pairs across all $\lambda$. In contrast, the production of large kinks from small kink collisions may be feasible only for small $\lambda$, where $m$ and $M$ are comparable, provided the initial velocity is sufficiently high.

\begin{figure*}[t!]
\centering
  \subfigure[]{\includegraphics[width=0.45\textwidth]{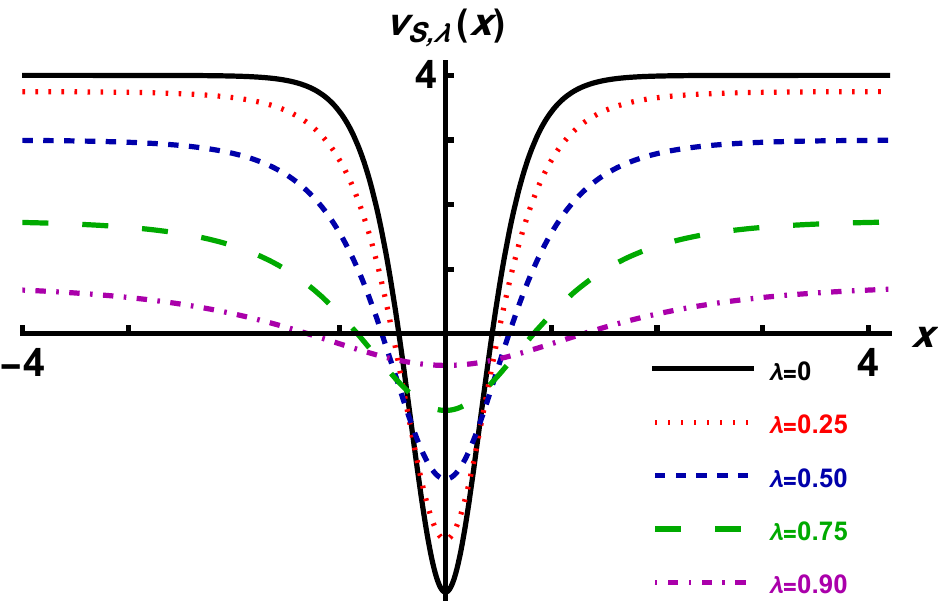}\label{fig:qmpS}}
  \hspace{1mm}
  \subfigure[]{\includegraphics[width=0.45\textwidth]{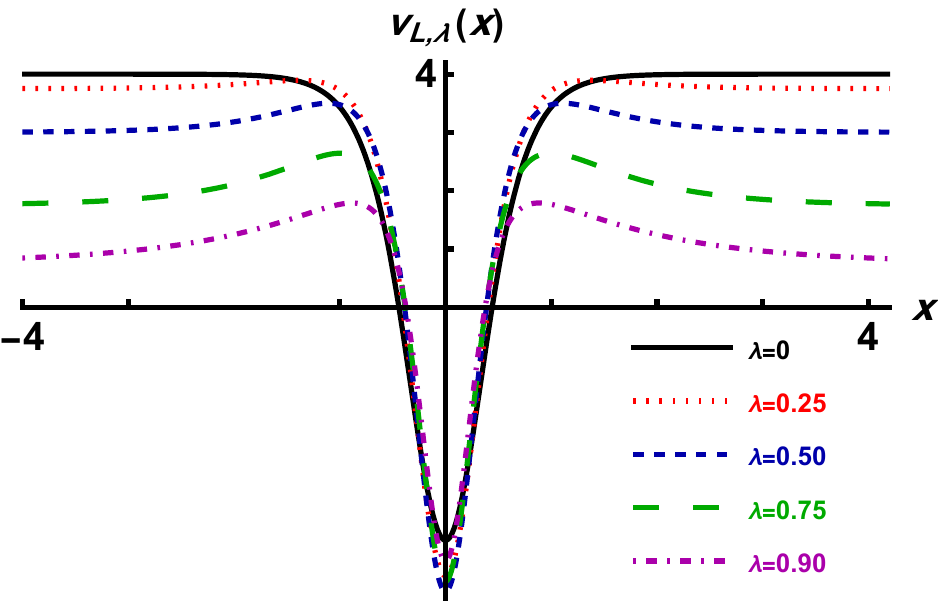}\label{fig:qmpL}}
\caption{(a) The stability potential $v_{S,\lambda}(x)$ for small kinks, as defined in Eq.~\eqref{eq:qmp_small}, for $\lambda = 0, 0.25, 0.50, 0.75, 0.90$. One sees that increasing $\lambda$, the potential well which has the Poeschl-Teller shape, becomes shallower and wider. (b) The stability potential $v_{L,\lambda}(x)$ for large kinks, as defined in Eq.~\eqref{eq:qmp_large}, for the same $\lambda$ values. For $\lambda$ from 0 to 0.54, the wells deepen and narrow, while for $\lambda > 0.54$, they become shallower and wider. We also note a change in the shape of the potential, from the reflectionless modified Poeschl-Teller to the volcano type as $\lambda$ increases towards unit.}
\label{fig:qmp}
\end{figure*}

To investigate the stability of the kink solutions and their internal modes, we introduce a time-dependent perturbation around the static kink, written as $\chi(x,t) = \chi_{K,\lambda}(x) + \sum_n \eta_n(x) \cos(\omega_n t)$, where $\chi_{K,\lambda}(x)$ represents either the small kink $\chi_{S,\lambda}(x)$ (Eq.~\eqref{smallk}) or the large kink $\chi_{L,\lambda}(x)$ (Eq.~\eqref{largek}), and $\eta_n(x)$ is a small perturbation. Substituting this into the equation of motion (Eq.~\eqref{eq:EOM}) and linearizing to first order in $\eta_n(x)$, we obtain a Schrödinger-like equation:
\begin{equation}\label{stabilityequation}
\left( -\frac{d^2}{dx^2} + v_{K,\lambda}(x) \right) \eta_n(x) = \omega_n^2 \eta_n(x),
\end{equation}
where the stability potential $v_{K,\lambda}(x)$ is defined as
\begin{eqnarray}\label{eq:qmp_small}
v_{S,\lambda}(x) = \left. \frac{d^2 U(\chi)}{d \chi^2} \right|_{\chi = \chi_{S,\lambda}(x)},
\end{eqnarray}
for small kinks, and
\begin{eqnarray}\label{eq:qmp_large}
v_{L,\lambda}(x) = \left. \frac{d^2 U(\chi)}{d \chi^2} \right|_{\chi = \chi_{L,\lambda}(x)},
\end{eqnarray}
for large kinks. The spectrum of Eq.~\eqref{stabilityequation} determines the vibrational modes of the kinks, with $\omega_n^2 > 0$ indicating stability. The stability potentials are illustrated in Figs.~\ref{fig:qmpS} and \ref{fig:qmpL} for small and large kinks, respectively. For small kinks (Fig.~\ref{fig:qmpS}), as $\lambda$ increases from 0 to 0.90, the potential wells become shallower (decreased depth) and wider (increased width), suggesting a reduction in the strength of internal modes. For large kinks (Fig.~\ref{fig:qmpL}), the behavior is more complex: as $\lambda$ increases from 0 to 0.5, the potential wells deepen (increased depth) and narrow (decreased width), potentially enhancing vibrational modes; however, for $\lambda > 0.5$, the wells become shallower and wider, weakening these modes. These variations in the stability potentials, driven by $\lambda$, directly influence the scattering dynamics, particularly the production of small kink-antikink pairs during large kink collisions. We further note that the stability potential for large kinks depicted in Fig. \ref{fig:qmpL} shows an interesting behavior, a change in its profile, from modified Poeschl-Teller to the volcano shape as $\lambda$ increases in the interval $\lambda\in (0,1)$. See Ref. \cite{Bazeia.EPJC.2017} for other details on this issue.

\section{Scattering}
\label{sec:Scattering}

Although the model lacks multisoliton solutions, the physical scenario of two kinks interactions can be effectively modeled and addressed through numerical methods. The kinks, as described in Eqs.~\eqref{largek} and \eqref{smallk}, enable us to adopt an initial setup consisting of a kink and an antikink, positioned at a significant distance apart and approaching each other. Such a setup approximates a solution to Eq.~\eqref{eq:EOM}, with deviations diminishing exponentially with the separation between the kink and antikink.

We conducted numerical simulations to explore kink-antikink scattering across various topological domains. To achieve this, we employed a numerical approach, utilizing a fourth-order spatial discretization scheme alongside the Stormer time integration technique. The discretized form of Eq.~\eqref{eq:EOM} used in our computations is:
\begin{eqnarray}
    \frac{d^2 \chi_{n}}{d t^2} - \frac{1}{h^2} (\chi_{n-1}-2\chi_{n}+\chi_{n+1}) +\nonumber\\
    +\frac{1}{12h^2}(\chi_{n-2}-4\chi_{n-1}+6\chi_{n}-4\chi_{n+1}+\chi_{n+2})+\nonumber\\
    + \frac1{{\text{dn}(\chi_{n} ,\lambda )^3}}\Bigl(\text{cn}(\chi_{n} ,\lambda ) \left(2 \text{cn}(\chi_{n} ,\lambda )^2+\lambda -1\right)\nonumber\\
    \left(-2 \text{dn}(\chi_{n} ,\lambda )^2+\lambda ^2+\lambda -2\right) \text{sn}(\chi_{n} ,\lambda )\Bigr) \!=\! 0, \nonumber
\end{eqnarray}
where $n=0,\pm1,\pm2,...$, and the grid consists of $N=14000$ nodes. The chosen time and space increments were $\tau=0.005$ and $h=0.025$, respectively. The starting arrangement in each simulation featured a kink and an antikink, initially placed at $x=\pm X_i^{}$ and traveling toward one another with speeds $v_i^{}$. For example, in the $(-a,a)$ topological sector for a kink-antikink encounter, the initial conditions for solving the equation of motion numerically were derived from the expression
\begin{eqnarray}
    \chi_{K\bar{K},\lambda}^{(-a,a)}(x,t) &=& \chi_{K,\lambda}^{(-a,a)}\left(\frac{x+X_i^{}-v_i^{}t}{\sqrt{1-v_i^2}}\right)+\nonumber\\
    &+& \chi_{\bar{K},\lambda}^{(-a,a)}\left(\frac{x-X_i^{}+v_i^{}t}{\sqrt{1-v_i^2}}\right) - a.\label{kantik}
\end{eqnarray}
We now proceed to examine the results of these numerical simulations.

\subsection{Small Kinks Scattering}\label{subsec:sks}

In this subsection, we investigate the scattering dynamics of small kinks in the system described by the potential $U_\lambda(\chi)$ (Eq.~\eqref{pot1}). The behavior of small kink-antikink pairs during collisions is analyzed for various values of the parameter $\lambda$ and initial velocities. We focus on the critical velocity for separation, the role of vibrational modes, the possibility of large kink creation, and resonance phenomena in the scattering process.

\begin{figure*}[t!]
\begin{center}
  \centering
  \subfigure[]{\includegraphics[width=0.32\textwidth]{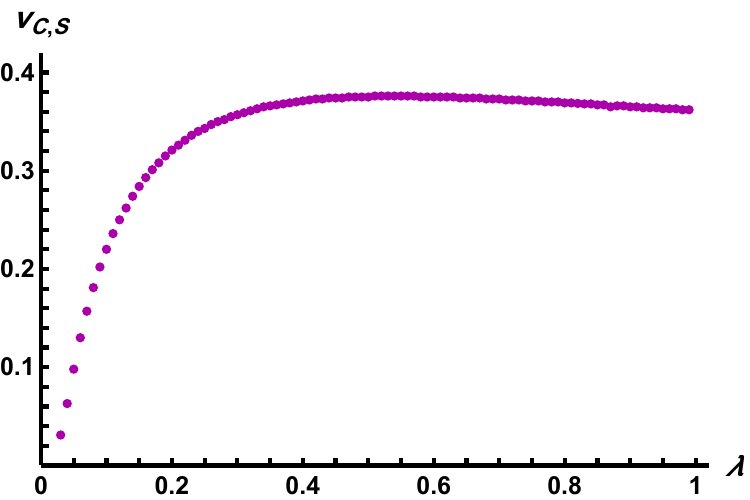}\label{fig:CriticalVelocitySmallKink}}
  \subfigure[]{\includegraphics[width=0.32\textwidth]{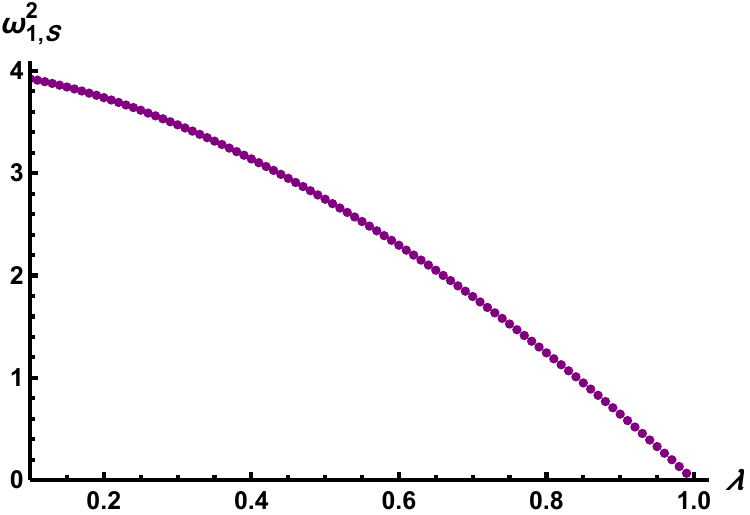}\label{fig:1stmodeSmallKinks}}
  \subfigure[]{\includegraphics[width=0.32\textwidth]{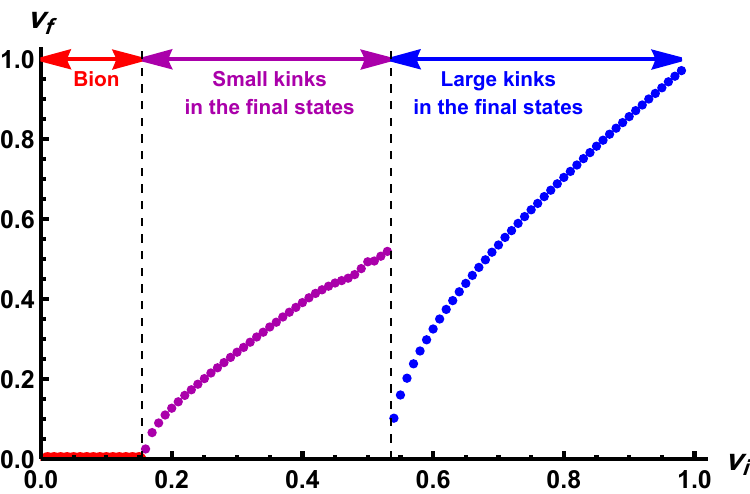}\label{fig:vfAsviLambda007}}
  \caption{(a) The critical velocity $v_c$ as a function of $\lambda$, showing a peak at $\lambda = 0.54$. (b) The first vibrational mode $\omega_1^2$ of small kinks as a function of $\lambda$, decreasing from 3.938 at $\lambda = 0.09$ to $0.067$ at $\lambda = 0.99$. (c) Final velocity $v_f$ versus initial velocity $v_i$ for $\lambda = 0.07$, showing bion formation ($v_i < 0.15$), small kink separation ($0.15 < v_i < 0.53$), and large kink creation ($v_i > 0.53$).}
  \label{fig:sks1}
\end{center}
\end{figure*}

Concerning kink collisions, Fig.~\ref{fig:sks1} presents an overview of the key characteristics of small kink scattering. In Fig.~\ref{fig:CriticalVelocitySmallKink}, the critical velocity $v_c$ for small kink-antikink scattering is plotted as a function of $\lambda$, with a step size of $\delta\lambda = 0.01$. The critical velocity $v_c$ represents the minimum initial velocity required for the kink and antikink to separate after collision. For $\lambda$ ranging from 0.03 to 0.54, $v_c$ increases from 0.031 to its maximum value of 0.376, indicating that higher initial velocities are needed for separation as $\lambda$ increases in this range. Beyond $\lambda = 0.54$, $v_c$ slightly decreases to 0.362 at $\lambda = 0.99$. This non-monotonic behavior suggests a complex interplay between the stability potential of small kinks (Fig.~\ref{fig:qmpS}) and the overall dynamics of the system.

The vibrational mode of small kinks also plays a crucial role in their scattering dynamics. In Fig.~\ref{fig:1stmodeSmallKinks}, the frequency of the first vibrational mode $\omega_1^2$ of small kinks is shown as a function of $\lambda$, with a step size of $\delta\lambda = 0.01$. Starting at $\lambda = 0.09$, $\omega_1^2$ takes its maximum value of 3.938, then decreases monotonically to 2.572 at $\lambda = 0.54$, and to approximately 0.5 at $\lambda = 0.99$. This decrease in $\omega_1^2$ corresponds to the shallowing and widening of the stability potential wells of small kinks (Fig.~\ref{fig:qmpS}), which reduces the ability of small kinks to store energy in their vibrational modes as $\lambda$ increases.

The scattering of small kinks is further explored in Fig.~\ref{fig:vfAsviLambda007}, which shows the final velocity $v_f$ as a function of the initial velocity $v_i$ for $\lambda = 0.07$. Three distinct regimes are observed. For $v_i$ from 0 to 0.15, $v_f = 0$ (red line), the result indicates that the kink and antikink form a bion — a bound state oscillating at the collision point with $n$ bounces. For $v_i$ between 0.15 and 0.53 (purple line), the small kinks separate after the collision, with $v_f$ increasing almost linearly with $v_i$, although $v_f < v_i$ occurs due to the energy loss in vibrational modes and radiation. For $v_i > 0.53$ (blue line), the collision results in the creation of a pair of large kinks. Since the mass of large kinks ($M$) is greater than that of small kinks ($m$), the final velocity of large kinks is lower than that of small kinks for the same $v_i$, consistent with energy conservation.

\begin{figure*}[t!]
\begin{center}
  \centering
  \subfigure[]{\includegraphics[width=0.45\textwidth]{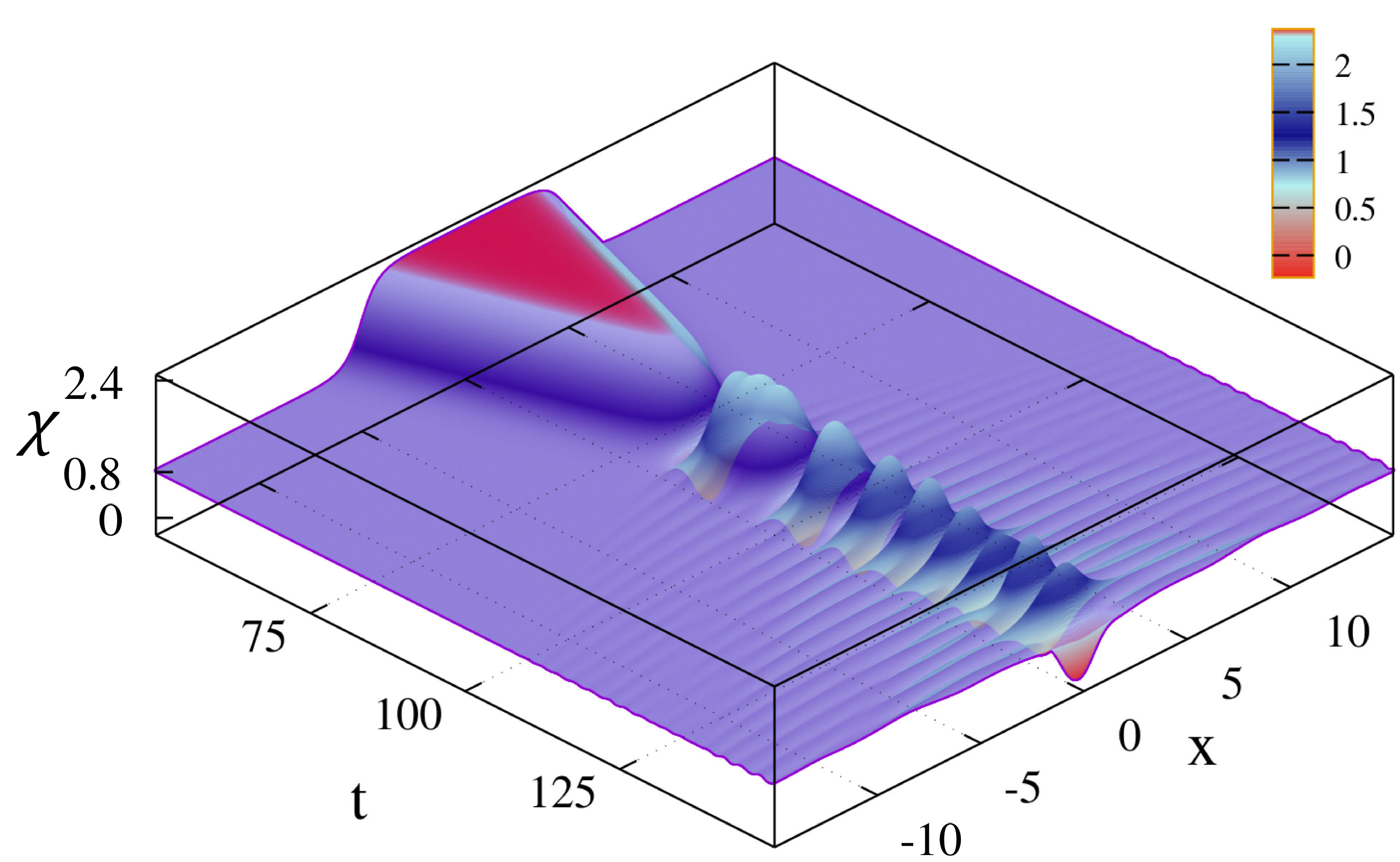}\label{fig:v01x10lambda07}}
  \subfigure[]{\includegraphics[width=0.45\textwidth]{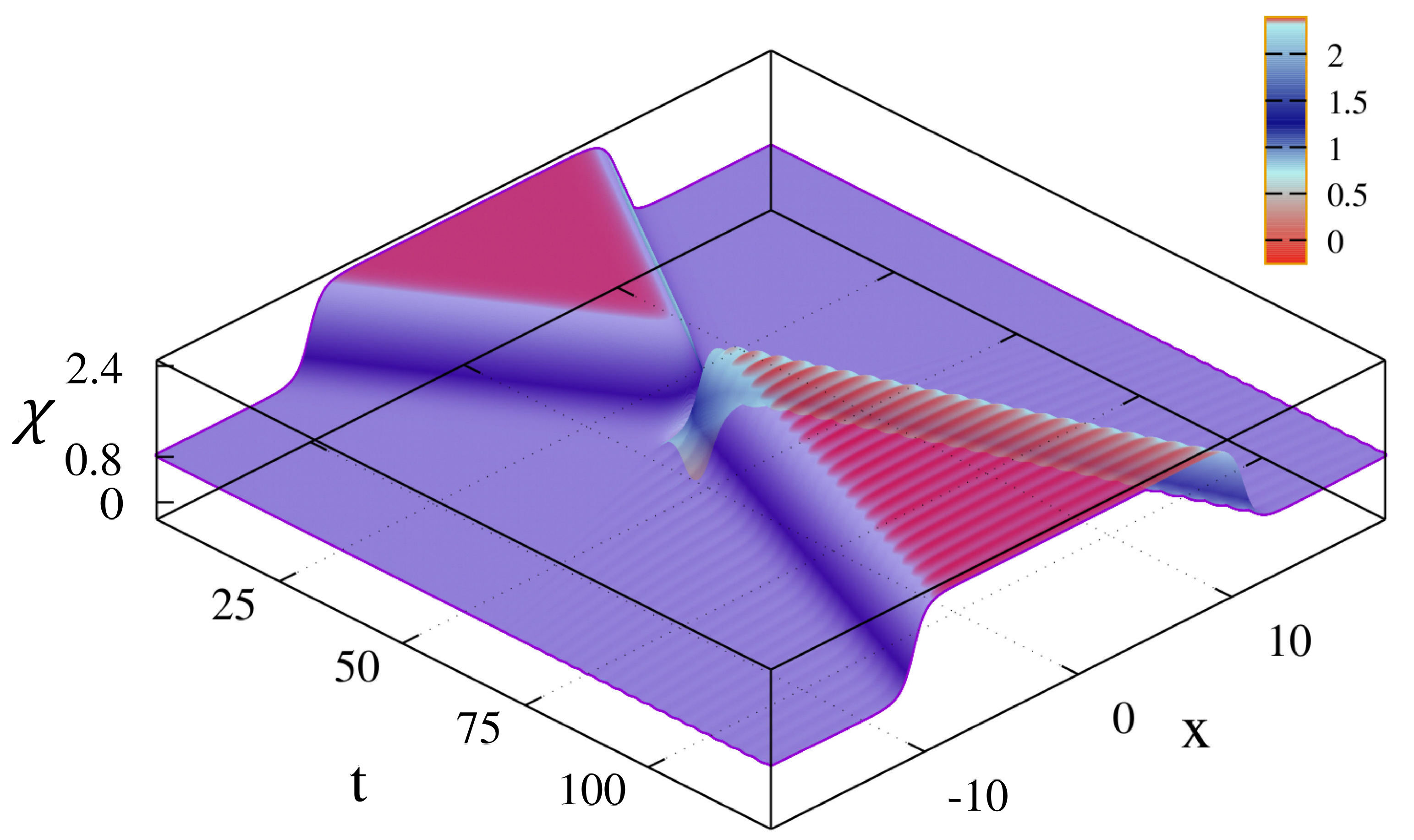}\label{fig:v02x10lambda07}}
  \\
  \subfigure[]{\includegraphics[width=0.45\textwidth]{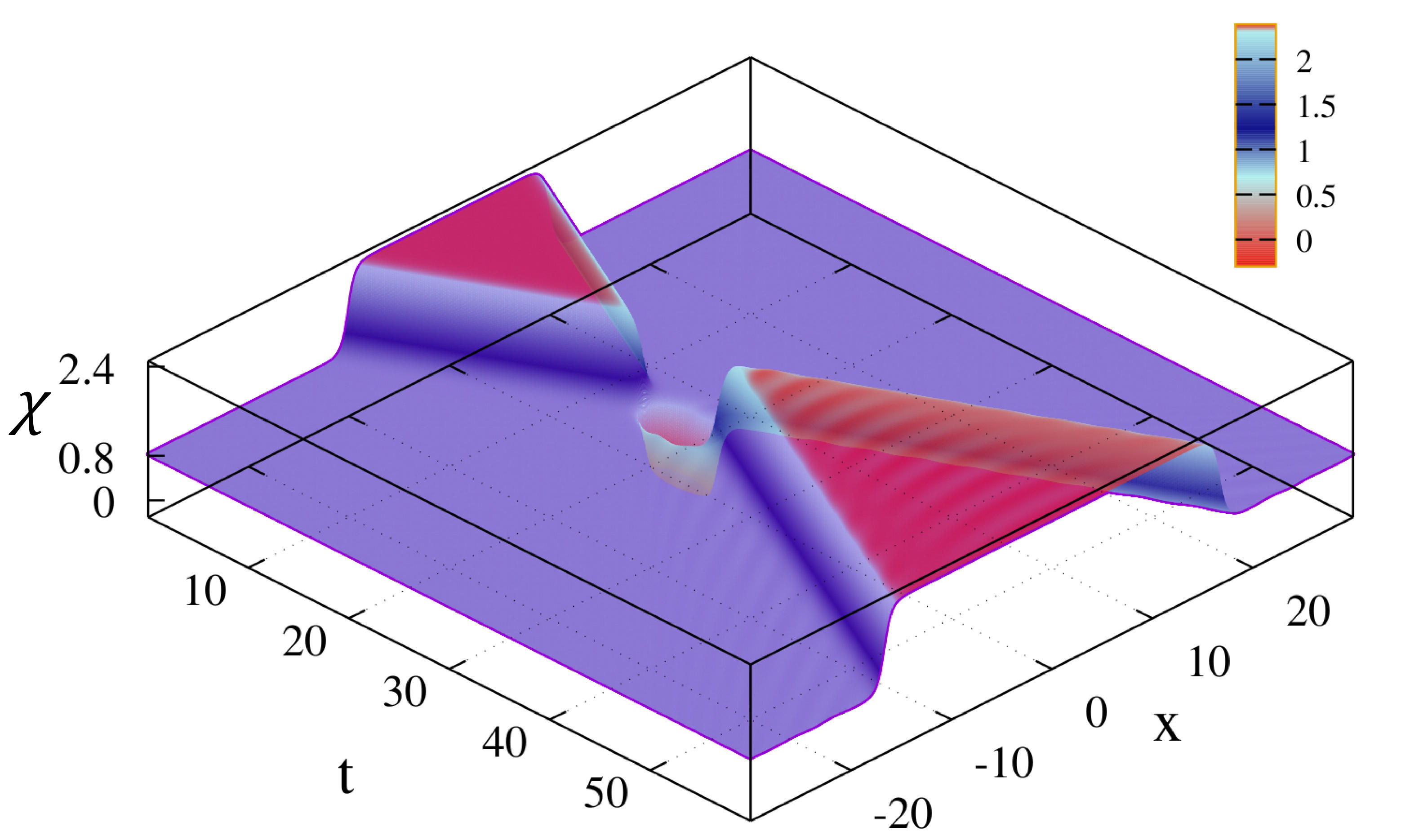}\label{fig:v05x10lambda07}}
  \subfigure[]{\includegraphics[width=0.45\textwidth]{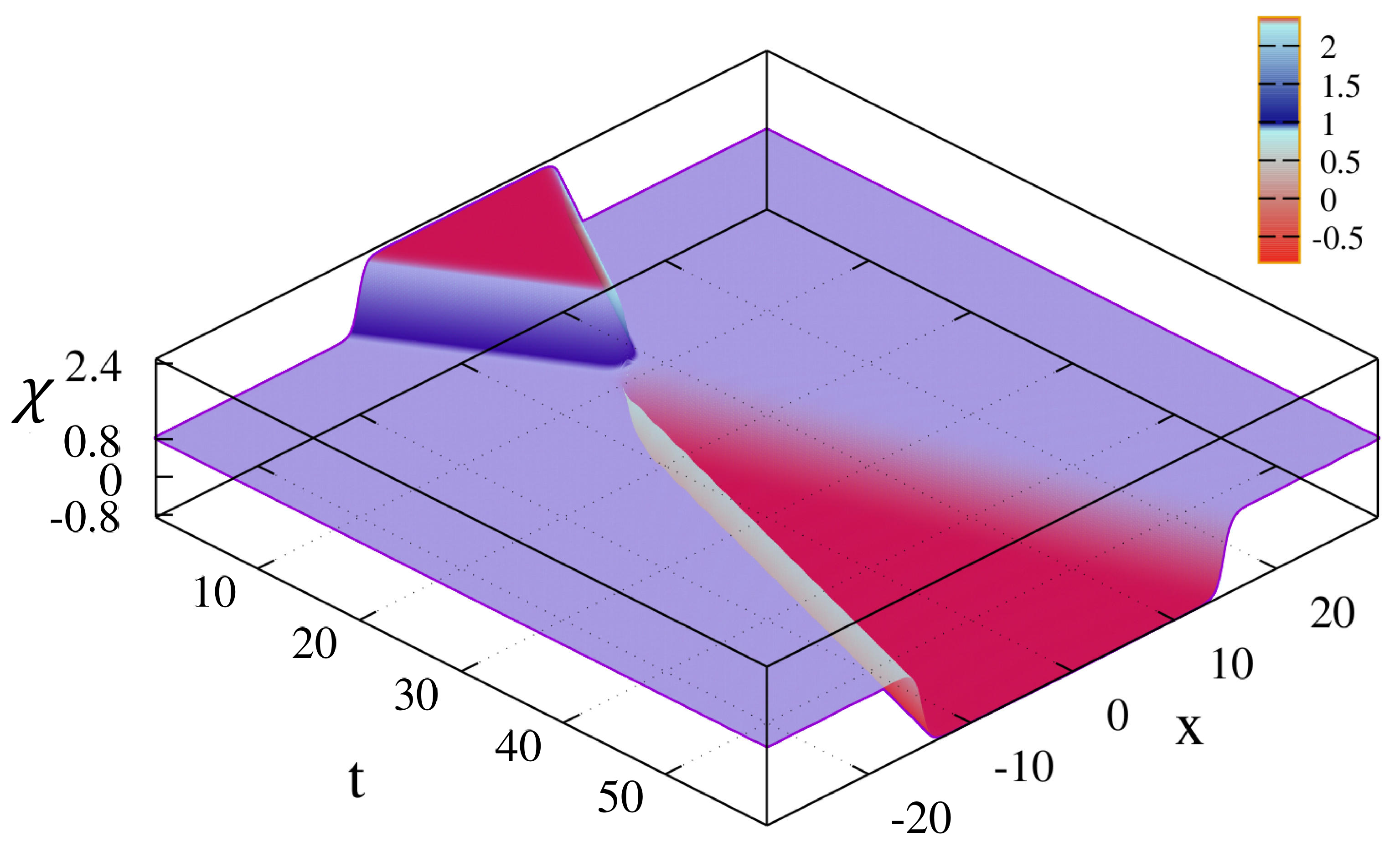}\label{fig:v06x10lambda07}}\vspace{0.3cm}
  \caption{Illustration of the small kinks scattering for $\lambda=0.07$ and different values of initial velocities, with initial positions $X_{0k}=-X_{0ak}=-10$ for all cases. (a) At $v_i = 0.1$, the kink and antikink form a bion. (b) At $v_i = 0.2$, the kink and antikink separate after collision. (c) At $v_i = 0.5$, separation occurs with a higher final velocity. (d) At $v_i = 0.6$, large kinks are created.}
  \label{fig:sks2}
\end{center}
\end{figure*}

To illustrate the scattering dynamics in more detail, Fig.~\ref{fig:sks2} shows the evolution of the field $\chi(x,t)$ for $\lambda = 0.07$ at different initial velocities, with initial positions $X_{0k} = -X_{0ak} = -10$, for kink $(k)$ and antikink $(ak)$. In Fig.~\ref{fig:v01x10lambda07}, for $v_{0k} = -v_{0ak} = 0.1$, the kink and antikink become trapped after collision, forming a bion that oscillates at $x \approx 0$ with $n$ bounces. This is consistent with Fig.~\ref{fig:vfAsviLambda007}, as $v_i = 0.1$ is below the critical velocity $v_c = 0.15$ for $\lambda = 0.07$. In Fig.~\ref{fig:v02x10lambda07}, with $v_{0k} = -v_{0ak} = 0.2$, the kink and antikink separate after collision, moving away symmetrically, as $v_i = 0.2$ exceeds $v_c$. Similarly, in Fig.~\ref{fig:v05x10lambda07}, for $v_{0k} = -v_{0ak} = 0.5$, the small kinks separate with a higher final velocity, still within the small kink regime ($v_i < 0.53$). Finally, in Fig.~\ref{fig:v06x10lambda07}, for $v_{0k} = -v_{0ak} = 0.6$, the collision leads to the creation of large kinks, consistent with the blue line in Fig.~\ref{fig:vfAsviLambda007}. The large kinks, having  greater mass, move with a lower final velocity compared to small kinks at the same initial velocity.

\begin{figure*}[t!]
\begin{center}
  \centering
  \subfigure[$\lambda=0.25$]{\includegraphics[width=0.45\textwidth]{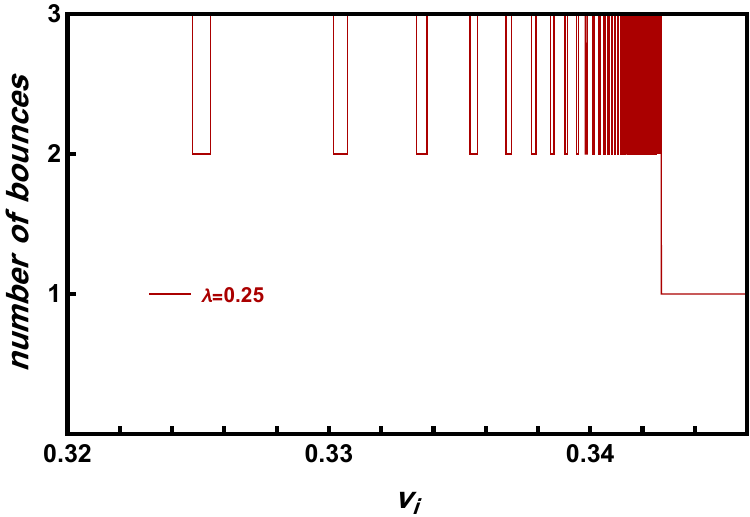}\label{fig:Nb_SK_landa_025}}
  \subfigure[$\lambda=0.25$]{\includegraphics[width=0.465\textwidth]{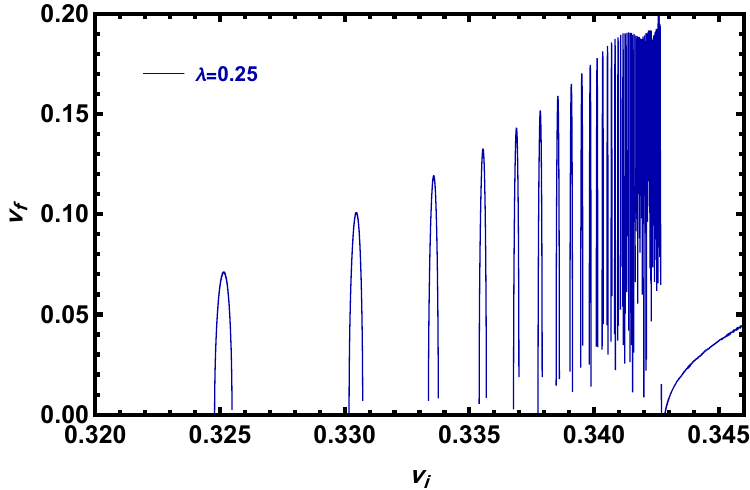}\label{fig:Vf_Vi_SK_landa_025}}
  \subfigure[$\lambda=0.50$]{\includegraphics[width=0.45\textwidth]{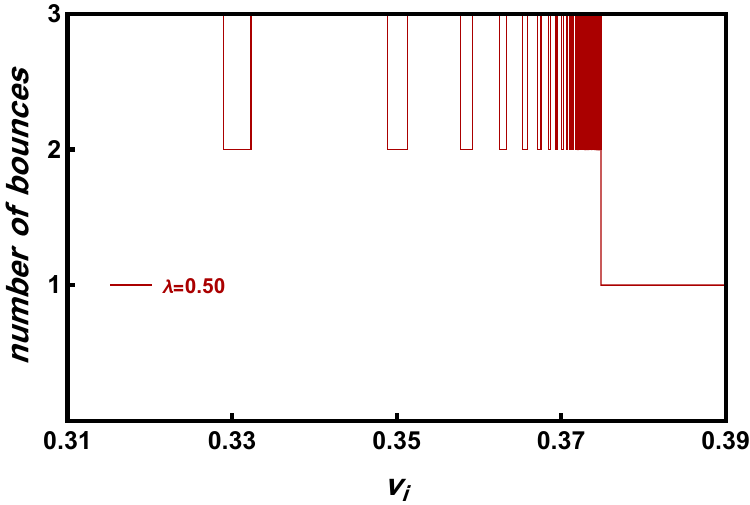}\label{fig:Nb_SK_landa_050}}
  \subfigure[$\lambda=0.50$]{\includegraphics[width=0.46\textwidth]{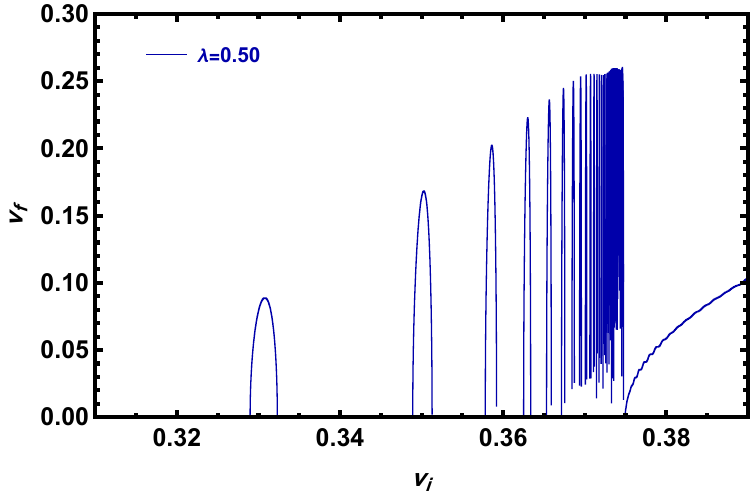}\label{fig:Vf_Vi_SK_landa_050}}
  \\
  \subfigure[$\lambda=0.75$]{\includegraphics[width=0.45\textwidth]{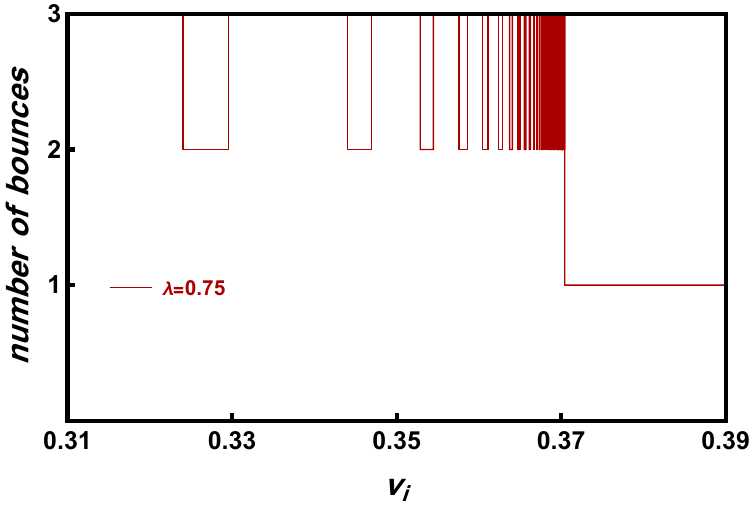}\label{fig:Nb_SK_landa_075}}
  \subfigure[$\lambda=0.75$]{\includegraphics[width=0.46\textwidth]{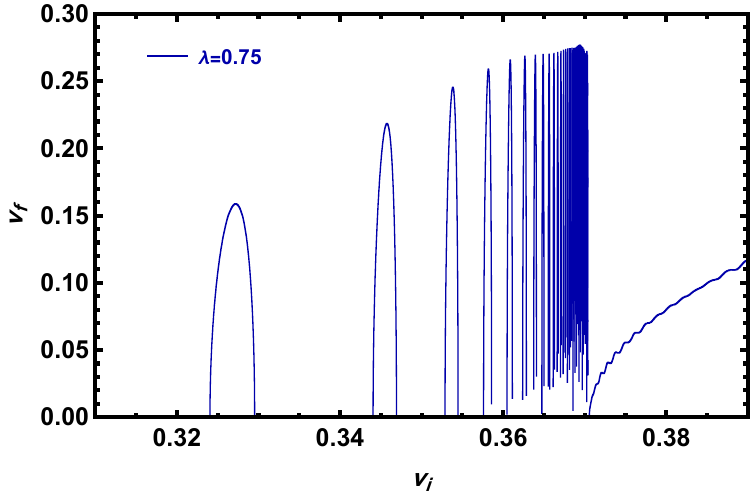}\label{fig:Vf_Vi_SK_landa_075}}
  \caption{Small kink-antikink collision, (a) and (b) correspond to simulations with parameter $\lambda=0.25$, (c) and (d) correspond to $\lambda=0.5$, while (e) and (f) correspond to $\lambda=0.75$. (a), (c), and (e) depict the number of bounces as a function of initial velocity and (b), (d), and (f) present the final velocity for cases with one and two bounces as a function of initial velocity. We are using $X_{0k}=-10$.}
  \label{fig:kaksmall}
\end{center}
\end{figure*}

The resonance phenomena in small kink-antikink collisions are explored in Fig.~\ref{fig:kaksmall}, which examines the number of bounces and final velocities for $\lambda = 0.25$, 0.5, and 0.75. In Fig.~\ref{fig:Nb_SK_landa_025}, for $\lambda = 0.25$, the number of bounces is shown as a function of $v_i$. Below $v_i = 0.324784$, the kinks form a bion, exhibiting $n$ bounces as an oscillatory bound state. For $v_i$ from 0.324784 to the critical velocity $v_c = 0.342718$, resonance windows are observed, with the number of bounces reaching $2$ or more, indicating that the kinks collide multiple times before separating due to energy exchange with their vibrational modes. The width of this resonance window is $0.017934$. For $v_i > 0.342718$, the number of bounces is typically $1$, as the kinks separate after a single collision. The corresponding final velocities are shown in Fig.~\ref{fig:Vf_Vi_SK_landa_025}. For $v_i$ from $0.324784$ to $0.342718$, where two-bounce or higher cases occur, $v_f$ varies between $0$ and $0.20$, reflecting the complex energy exchange in vibrational modes and radiation. Notably, in cases where $v_f$ reaches higher values (e.g., up to $0.20$), energy is transferred from the internal vibrational modes of the kinks to their translational modes, leading to a higher final velocity. For $v_i > 0.342718$, where one-bounce cases dominate, $v_f$ increases with $v_i$, but remains lower than the maximum values observed in the two-bounce regime, as less energy is exchanged in multiple collisions. This is an interesting result, since it indicates that the final velocity in the two-bounce regime can be higher than in the one-bounce regime, due to the resonant energy transfer from internal to translational modes in the former case.

For $\lambda = 0.5$, Fig.~\ref{fig:Nb_SK_landa_050} shows that the critical velocity increases to $v_c = 0.374851$, consistent with Fig.~\ref{fig:CriticalVelocitySmallKink}. Below $v_i = 0.328967$, the kinks form a bion with $n$ bounces. Resonance windows are still present, with two bounces or more observed for $v_i$ from 0.328967 to 0.374851. The width of the resonance window increases to $0.045884$, indicating a broader range of initial velocities where resonance occurs compared to $\lambda = 0.25$. For $v_i > 0.374851$, the kinks separate after one bounce. The final velocities in Fig.~\ref{fig:Vf_Vi_SK_landa_050} follow a similar trend to $\lambda = 0.25$, with $v_f$ for two-bounce cases being higher than for one-bounce cases due to energy transfer from internal to translational modes, and $v_f$ increasing with $v_i$ for one-bounce cases. Finally, for $\lambda = 0.75$, Fig.~\ref{fig:Nb_SK_landa_075} indicates a critical velocity of $v_c = 0.370429$, which is slightly lower than that at $\lambda = 0.5$, consistent with the trend observed in Fig.~\ref{fig:CriticalVelocitySmallKink}. Below $v_i = 0.318815$, a bion forms with $n$ bounces. Resonance windows are nearly absent, with two bounces or more observed only in a narrow range of $v_i$ from 0.318815 to 0.370429. The width of this resonance window is $0.051614$, which is slightly larger than that at $\lambda = 0.5$. For $v_i > 0.370429$, the kinks separate after one bounce. The final velocities in Fig.~\ref{fig:Vf_Vi_SK_landa_075} reflect this simpler dynamics, showing a increase in $v_f$ with $v_i$.

The observed trends in resonance phenomena are directly related to the vibrational mode of small kinks. As $\lambda$ increases, the frequency $\omega_1^2$ decreases (Fig.~\ref{fig:1stmodeSmallKinks}), reducing the ability of small kinks to store energy in their vibrational modes. Consequently, resonance windows become less prominent at higher $\lambda$ values, leading to simpler scattering dynamics. The width of the resonance windows, however, shows a non-monotonic trend: it increases from 0.017934 at $\lambda = 0.25$ to 0.045884 at $\lambda = 0.5$, but then slightly increases further to 0.051614 at $\lambda = 0.75$, despite the near absence of resonance windows at this $\lambda$. The critical velocity $v_c$ increases from 0.342718 at $\lambda = 0.25$ to 0.374851 at $\lambda = 0.5$, reflecting the increasing difficulty of separation due to changes in the potential $U_\lambda(\chi)$. However, $v_c$ slightly decreases to 0.370429 at $\lambda = 0.75$, consistent with the trend in Fig.~\ref{fig:CriticalVelocitySmallKink}, possibly due to the weakening of vibrational modes. Additionally, the absence of vibrational modes in large kinks, as determined earlier, explains why large kink creation is only observed at low $\lambda$ values (e.g., $\lambda = 0.07$), where the mass difference $M - m$ is smaller (Fig.~\ref{fig:mass}), making the transition energetically feasible.

\subsection{Large Kinks Scattering}\label{subsec:lks}

In this subsection, we explore the scattering dynamics of large kinks in the system described by the potential $U_\lambda(\chi)$ shown in Eq.~\eqref{pot1}. Unlike small kinks, large kinks lack vibrational modes, which significantly influences their scattering behavior. Here, we focus on the outcomes of large kink-antikink collisions, particularly the creation of small kinks, for various values of $\lambda$ and initial velocities.

\begin{figure*}[t!]
\begin{center}
  \centering
  \subfigure[]{\includegraphics[width=0.45\textwidth]{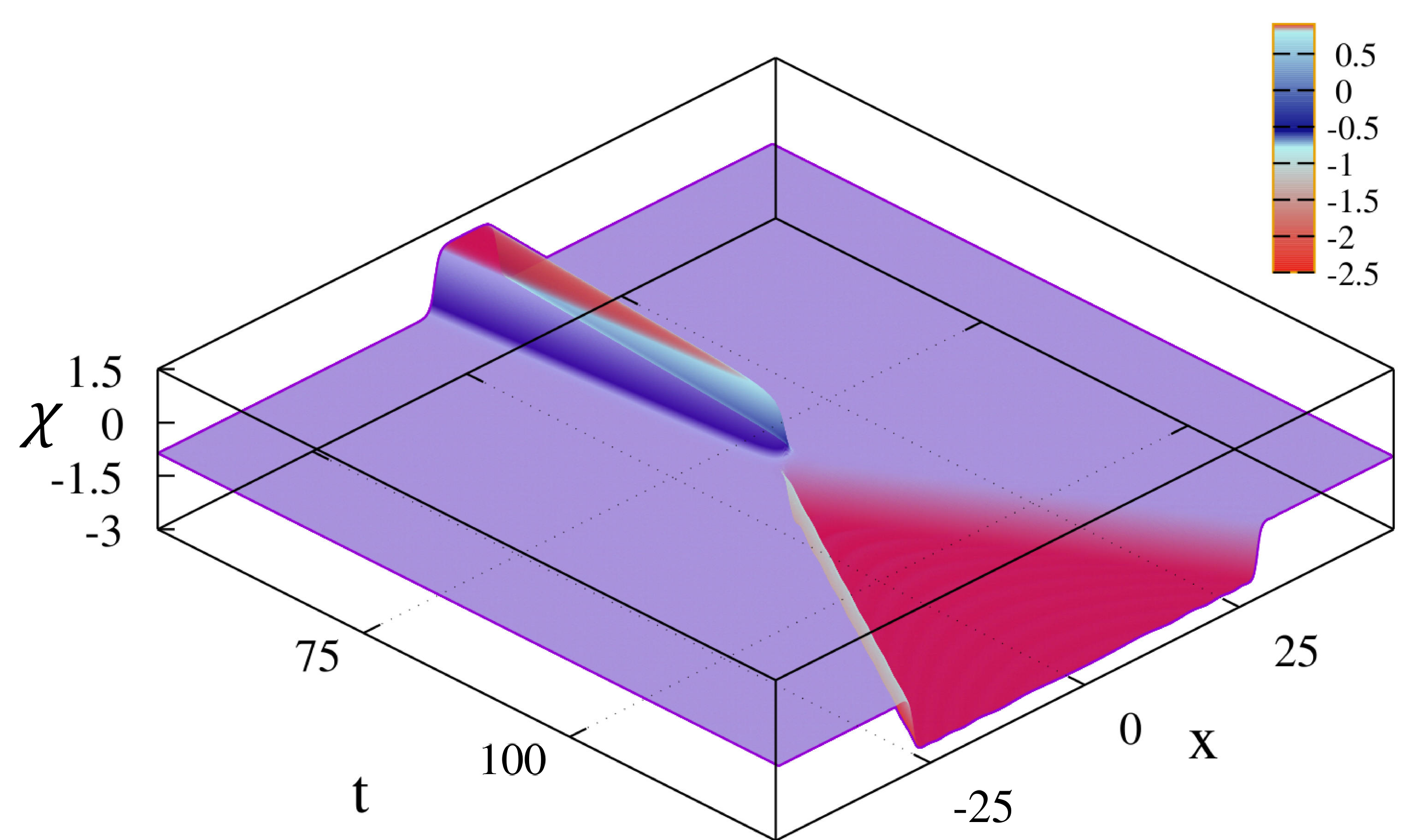}\label{fig:v01x10lambda02LK}}
  \subfigure[]{\includegraphics[width=0.45\textwidth]{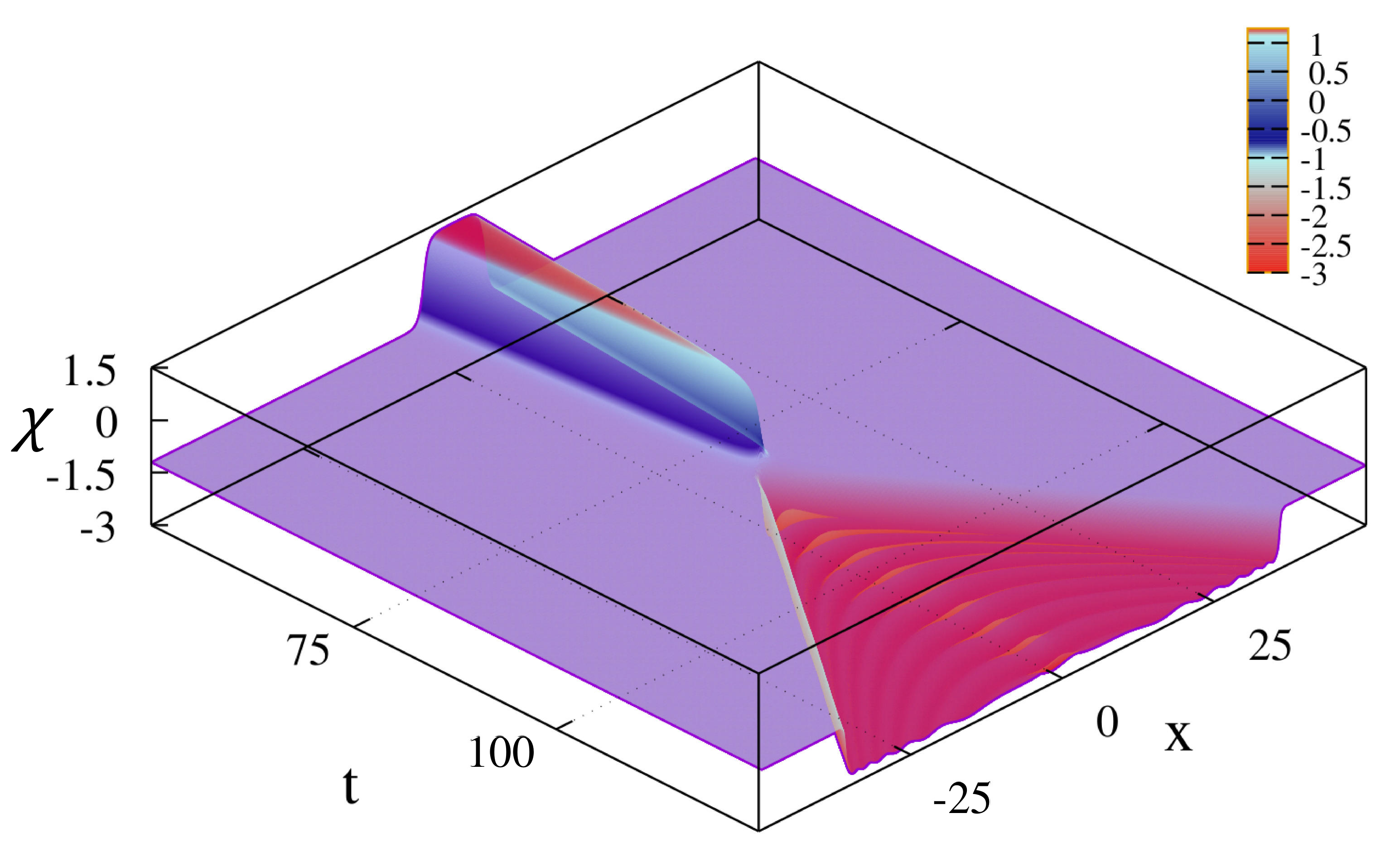}\label{fig:v01x10lambda06LK}}
  \\
  \subfigure[]{\includegraphics[width=0.45\textwidth]{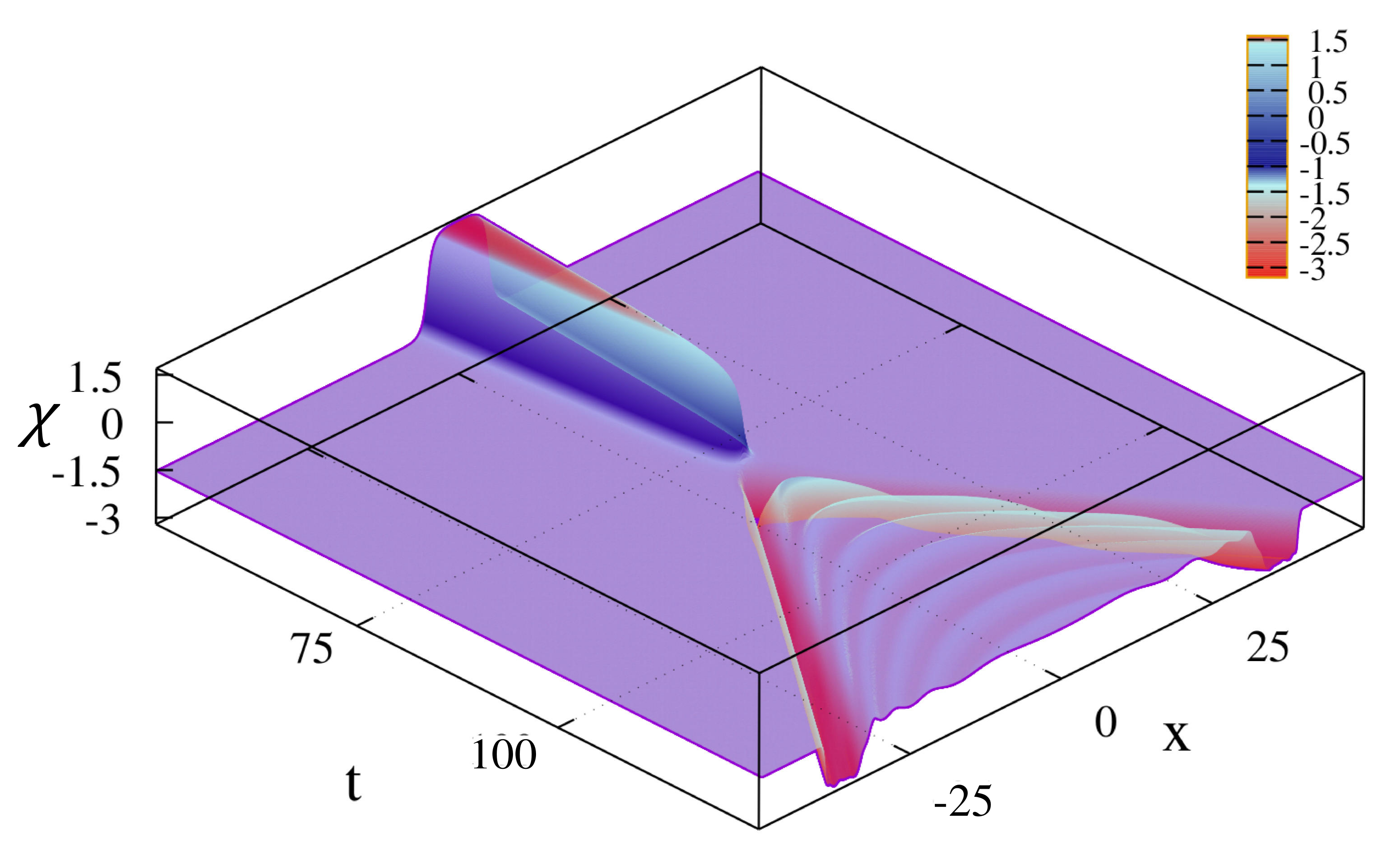}\label{fig:v01x10lambda08LK}}
  \subfigure[]{\includegraphics[width=0.45\textwidth]{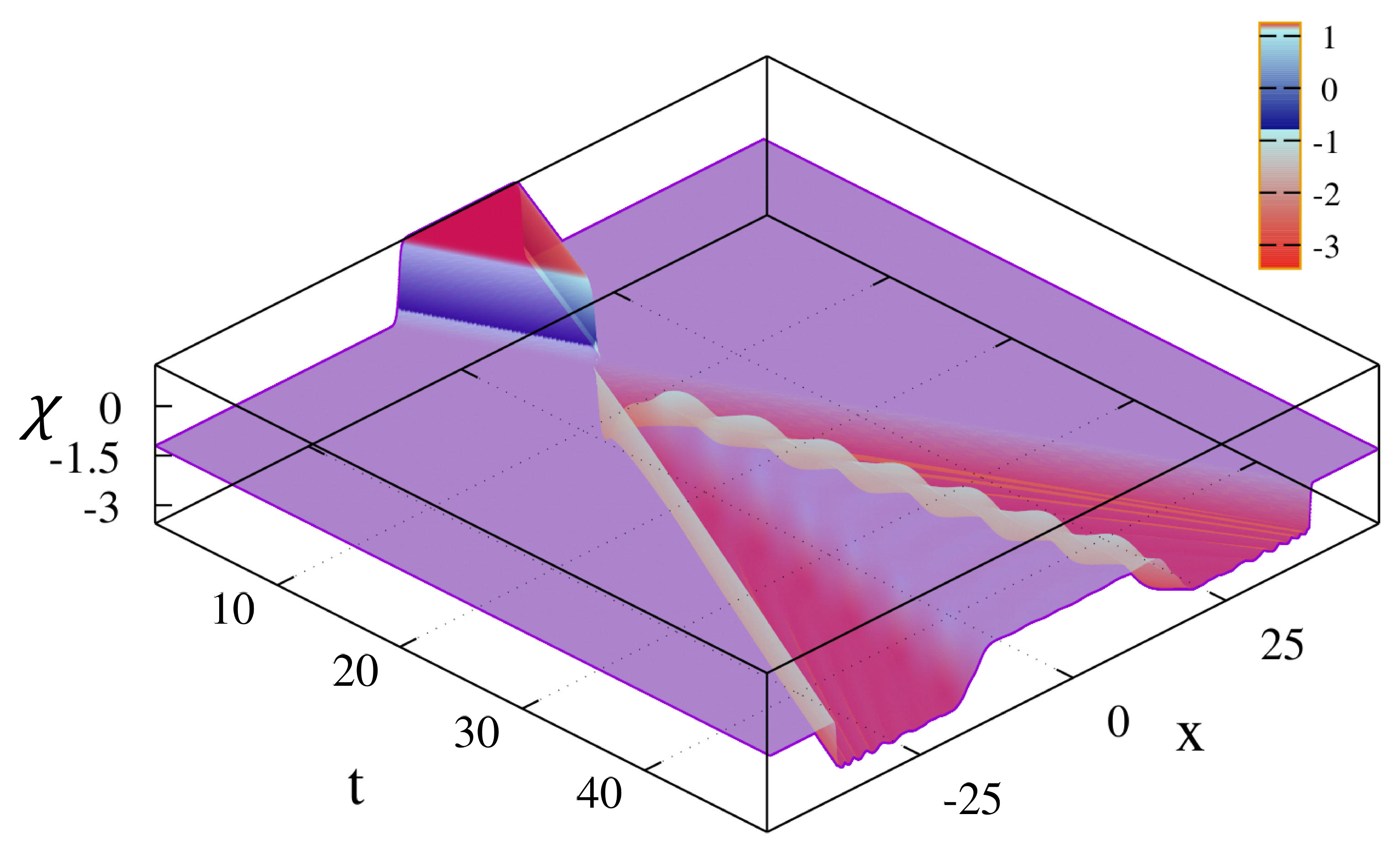}\label{fig:v09x10lambda06LK}}
  \\
  \subfigure[]{\includegraphics[width=0.45\textwidth]{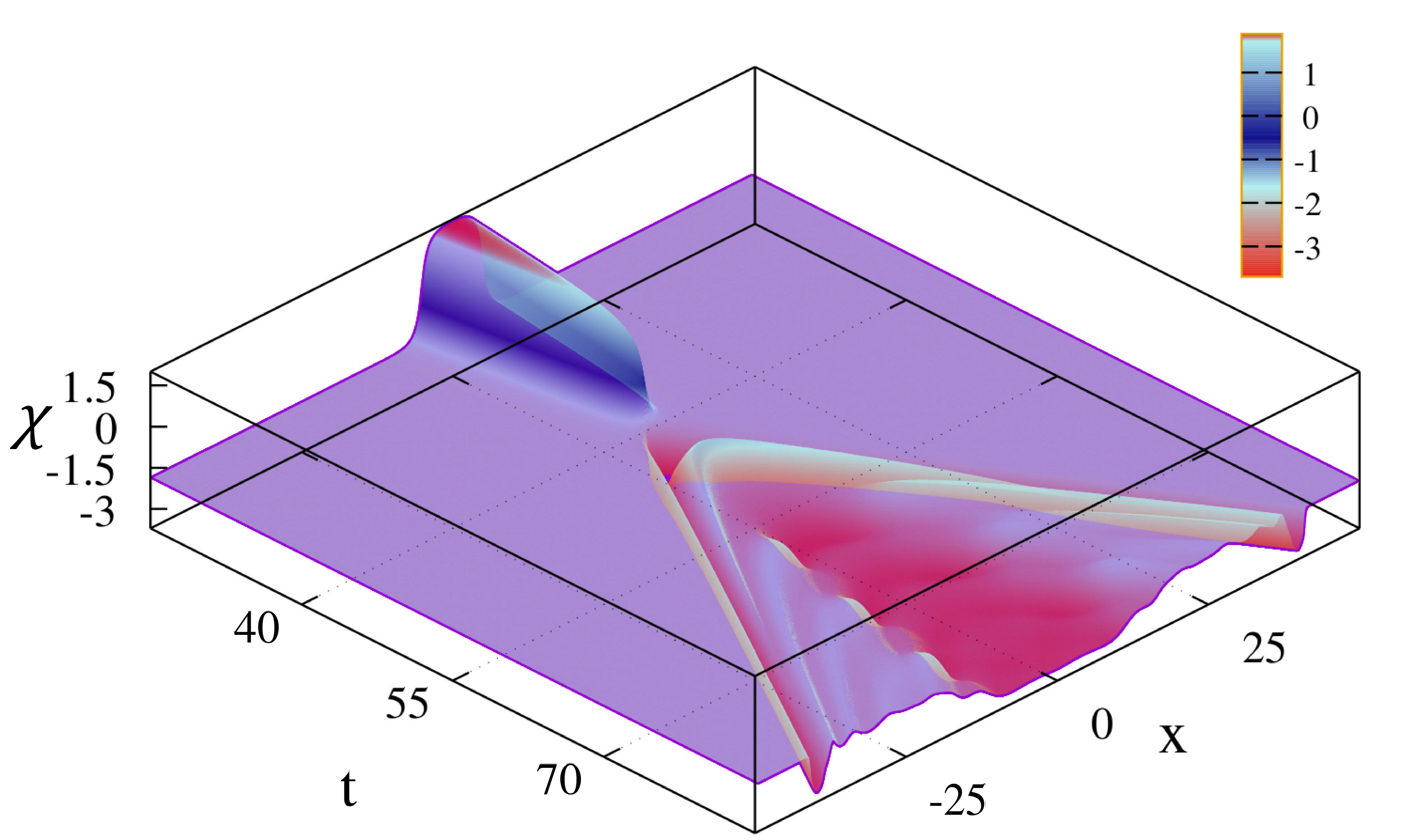}\label{fig:v02x10lambda09LK}}
  \subfigure[]{\includegraphics[width=0.45\textwidth]{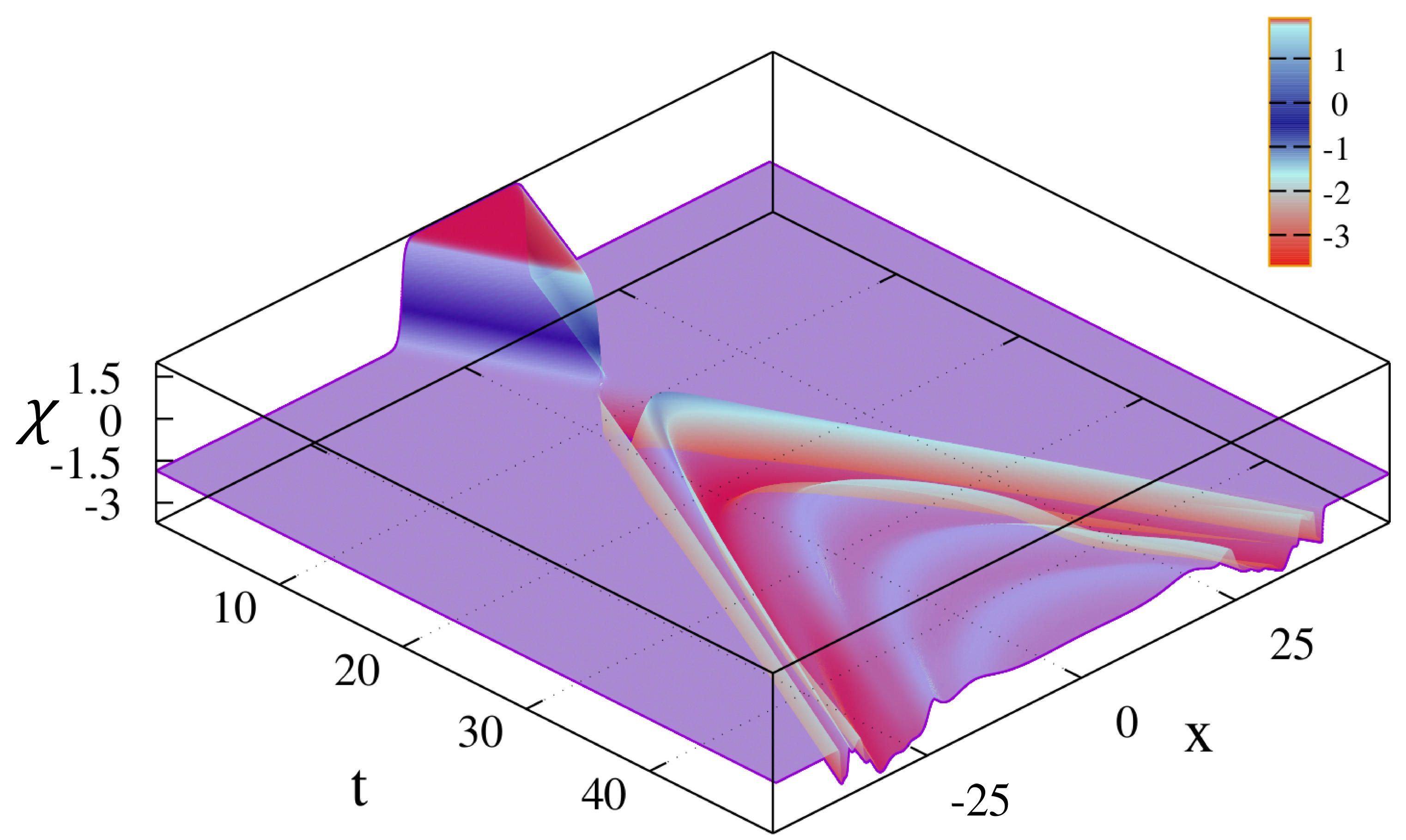}\label{fig:v09x10lambda09LK}}
  \caption{Large kinks scattering for different values of $\lambda$ and initial velocities, with initial positions $X_{0k}=-X_{0ak}=-10$ for all cases. (a) At $\lambda = 0.2$ and $v_{0k} = 0.1$, two small kinks are created. (b) At $\lambda = 0.6$ and $v_{0k} = 0.1$, two small kinks with higher velocities are produced. (c) At $\lambda = 0.8$ and $v_{0k} = 0.1$, two pairs of small kinks are created. (d) At $\lambda = 0.6$ and $v_{0k} = 0.9$, two pairs of small kinks are produced. (e) At $\lambda = 0.9$ and $v_{0k} = 0.2$, three pairs of small kinks are created. (f) At $\lambda = 0.9$ and $v_{0k} = 0.9$, four pairs of small kinks are produced.}
  \label{fig:lks}
\end{center}
\end{figure*}

Fig.~\ref{fig:lks} illustrates the scattering of large kinks for different values of $\lambda$ and initial velocities, with initial positions $X_{0k} = -X_{0ak} = -10$ in all cases. In Fig.~\ref{fig:v01x10lambda02LK}, for $\lambda = 0.2$ and $v_{0k} = -v_{0ak} = 0.1$, the collision results in the creation of two small kinks, which move away symmetrically after the interaction. Increasing $\lambda$ to 0.6 with the same initial velocity ($v_{0k} = -v_{0ak} = 0.1$), as shown in Fig.~\ref{fig:v01x10lambda06LK}, also leads to the creation of two small kinks, but with higher velocities, indicating that a larger $\lambda$ enhances the energy transfer to the resulting small kinks. For $\lambda = 0.8$ and $v_{0k} = -v_{0ak} = 0.1$ (Fig.~\ref{fig:v01x10lambda08LK}), the collision produces two pairs of small kinks, suggesting that higher $\lambda$ values facilitate the creation of additional small kinks due to changes in the potential $U_\lambda(\chi)$. A similar outcome is observed in Fig.~\ref{fig:v09x10lambda06LK} for $\lambda = 0.6$ but with a higher initial velocity ($v_{0k} = -v_{0ak} = 0.9$), where two pairs of small kinks are created, highlighting the role of initial velocity in increasing the number of small kinks produced. For $\lambda = 0.9$ and $v_{0k} = -v_{0ak} = 0.2$ (Fig.~\ref{fig:v02x10lambda09LK}), the collision results in the creation of three pairs of small kinks, and increasing the initial velocity to $v_{0k} = -v_{0ak} = 0.9$ at the same $\lambda$ (Fig.~\ref{fig:v09x10lambda09LK}) leads to the creation of four pairs of small kinks. As expected, the results depicted in Figure 6 show that both higher $\lambda$ values and larger initial velocities directly contribute to the production of more small kinks, as the increased energy available in the collision can overcome the energy barriers required to create additional small kinks.

The creation of small kinks in large kink-antikink collisions can be understood through energy conservation. Large kinks, having a greater mass ($M > m$, where $m$ is the mass of a small kink), possess more kinetic energy for the same initial velocity. During the collision, this kinetic energy is partially converted into the potential energy required to create small kinks, with the number of small kinks produced depending on the available energy, which is influenced by both $\lambda$ and the initial velocity. The absence of vibrational modes in large kinks, as noted earlier, means that the energy exchange during the collision is primarily between translational modes and the potential energy of the system, facilitating the creation of small kinks rather than energy storage in internal modes. A quantitative investigation dealing with the number of small kinks created in terms of $\lambda$ and initial velocity $v_i$ of the kink and antikink system described in Eq. \eqref{kantik} is also of current interest, since it may be further connected with quantum effects for kink creation, as recently investigated in Ref. \cite{vasha}. This motivation requires adding other fields, which is out of the scope of the present work and will be left to a future investigation.

\section{Conclusion}\label{sec:conclusion}

In this study, we investigated the scattering dynamics of kinks in a system governed by the potential $U_\lambda(\chi)$ shown in Eq.~\eqref{pot1}, focusing on both small and large kinks. Our analysis revealed distinct behaviors in the scattering processes of small and large kinks, driven by the interplay of the system parameter $\lambda$, initial velocities, and the intrinsic properties of the kinks.

For small kinks, we observed a rich variety of scattering outcomes, including bion formation, separation after collision, and the creation of large kinks, depending on the initial velocity $v_i$ and the parameter $\lambda$. The critical velocity $v_c$ for separation exhibits a non-monotonic dependence on $\lambda$, peaking at $\lambda = 0.54$ with $v_c = 0.376$, and slightly decreasing to 0.362 at $\lambda = 0.99$. This behavior reflects the complex influence of the potential $U_\lambda(\chi)$ on the stability and dynamics of small kinks. Additionally, the vibrational modes of small kinks play a crucial role in their scattering dynamics. The frequency of the first vibrational mode $\omega_1^2$ decreases from 3.938 at $\lambda = 0.09$ to 0.067 at $\lambda = 0.99$, reducing the ability of small kinks to store energy in their internal modes as $\lambda$ increases. This leads to a diminishing prominence of resonance windows at higher $\lambda$ values, with the width of resonance windows showing a non-monotonic trend: increasing from 0.017934 at $\lambda = 0.25$ to 0.045884 at $\lambda = 0.5$, and slightly further to 0.051614 at $\lambda = 0.75$. The creation of large kinks from small kink collisions is energetically favorable at low $\lambda$ values (e.g., $\lambda = 0.07$), where the mass difference $M - m$ is smaller, facilitating the transition.

In contrast, large kinks exhibit simpler scattering dynamics due to the absence of vibrational modes. Their collisions consistently result in the creation of small kinks, with the number of small kink pairs produced increasing with both $\lambda$ and initial velocity. For example, at $\lambda = 0.2$ and $v_{0k} = -v_{0ak} = 0.1$, two small kinks are created, while at $\lambda = 0.9$ and $v_{0k} = -v_{0ak} = 0.9$, four pairs of small kinks are produced. This trend underscores the role of energy conservation, where the greater mass of large kinks ($M > m$) provides more kinetic energy that is converted into the potential energy required to create small kinks during collisions.

Our findings highlight the significant influence of the parameter $\lambda$ on the scattering dynamics of kinks, mediating the transition between different kink scattering regimes and influencing the production of new kinks. The absence of vibrational modes in large kinks simplifies their dynamics, while the presence of such modes in small kinks introduces complex phenomena like resonance and bion formation. These results contribute to a deeper understanding of kink interactions in nonlinear systems and may have implications for applications in fields such as condensed matter physics, where topological defects such as kinks may play a critical role. It is well known that magnetic materials at the nanometric scale may produce localized magnetic configurations in the form of kinks or domain walls that are of current interest in applications of practical use, such as the implementation of digital devices \cite{MM01}. In the case of kinks that change their internal profile, it is of interest to recall the investigation developed before in \cite{MM02}, where the authors studied the possibility that in a nanowire, two distinct types of kinks (known in condensed matter as the Boch wall and the Néel wall) can switch their internal profile from Bloch to Néel wall under modification of the width and thickness of the nanowire. Future studies could explore the effects of additional parameters, such as external fields or temperature, on kink scattering, as well as the potential for experimental verification of these phenomena in physical systems.

\section*{Acknowledgments}

AMM thanks Islamic Azad University Quchan branch for the grant, and DB thanks the Conselho Nacional de Desenvolvimento Científico e Tecnológico (CNPq), grants Nos. 402830/2023-7 and 303469/2019-6 for partial financial support.


\begin{thebibliography}{99}

\bibitem{Rajaraman.book.1982}
R.~Rajaraman,
{\it Solitons and Instantons: An Introduction to Solitons and Instantons in Quantum Field Theory},
North-Holland (1982).

\bibitem{Vilenkin.book.2000}
A.~Vilenkin, and E.~P.~S.~Shellard,
{\it Cosmic Strings and Other Topological Defects},
Cambridge University Press (2000).

\bibitem{Manton.book.2004}
N.~Manton, and P.~Sutcliffe,
{\it Topological Solitons},
Cambridge University Press, Cambridge University Press (2004).

\bibitem{Shnir.book.2018}
Y.~M.~Shnir, {\it Topological and Non-Topological Solitons in Scalar Field Theories}\href{https://doi.org/10.1017/9781108555623},
 Cambridge University Press (2018). 

\bibitem{Bishop.PhysD.1980}
A.~R.~Bishop, J.~A.~Krumhansl, and S.~E.~Trullinger,
{\it Solitons in condensed matter: A paradigm},
\href{https://doi.org/10.1016/0167-2789(80)90003-2}{{\it{Physica  {\bf D}}} {\bf 1} (1980) 1 . }

 \bibitem{Optics2}
 S.~V.~Suchkov, A.~A.~Sukhorukov,  J.~Huang,  S.~V.~Dmitriev,  C.~Lee, and Yu.~S.~Kivshar, {\it Nonlinear switching and solitons in PT-symmetric photonic systems, }\href{ https://doi.org/10.1002/lpor.201500227} {{\it Laser Photonics Rev.} {\bf 10} (2016) 2, 177-213. }

\bibitem{Dmitriev.Nonlinearity.2000}
A.~E.~Miroshnichenko, S.~V.~Dmitriev, A.~A.~Vasiliev and T.~Shigenari, {\it Inelastic three-soliton collisions in a weakly discrete sine-Gordon system, }\href{ https://doi.org/10.1088/0951-7715/13/3/318} {{\it Nonlinearity.} {\bf 13} (2000) 837. }

\bibitem{Dmitriev.PRE.2008}
S.~V.~Dmitriev, P.~G.~Kevrekidis, and Y.~S.~Kivshar, {\it Radiationless energy exchange in three-soliton collisions, }\href{ https://doi.org/10.1103/PhysRevE.78.046604} {{\it Phys.\ Rev.\ {\bf E}} {\bf 78} (2008) 046604. }

\bibitem{Moradi.EPJB.2018}
A.~Moradi~Marjaneh, A.~Askari, D.~Saadatmand, and S.V.~Dmitriev,
{\it Extreme values of elastic strain and energy in sine-Gordon multi-kink collisions},
\href{https://doi.org/10.1140/epjb/e2017-80406-y}{{\it Eur.\ Phys.\ J.\ {\bf B}} {\bf 91} (2018) 22}
[\href{https://arxiv.org/abs/1710.10159}{\tt arXiv:1710.10159}].

\bibitem{Makhankov.PhysRep.1978}
V.G. Makhankov,
{\it Dynamics of classical solitons (in non-integrable systems)},
\href{https://doi.org/10.1016/0370-1573(78)90074-1}{{\it Phys.\ Rep.\ } {\bf 35} (1978) 1}.

\bibitem{Campbell.PhysD.1983}
D.~K.~Campbell, J.~F.~chonfeld, C.~A.~Wingate,
{\it Resonance structure in kink-antikink interactions in $\varphi^4$ theory},
\href{https://doi.org/10.1016/0167-2789(83)90289-0}{{\it{Physica  {\bf D}}} {\bf 9} (1980) 1-2 . }

\bibitem{Moradi.CNSNS.2017}
A.~Moradi~Marjaneh, D.~Saadatmand, K.~Zhou, S.V.~Dmitriev, and M.E.~Zomorrodian,
{\it High energy density in the collision of $N$ kinks in the $\phi^4$ model},
\href{https://doi.org/10.1016/j.cnsns.2017.01.022}{{\it Commun.\ Nonlinear Sci.\ Numer.\ Simulat.} {\bf 49} (2017) 30}
[\href{https://arxiv.org/abs/1605.09767}{\tt arXiv:1605.09767}].

\bibitem{Askari.CSF.2020}
A.~Askari, A.~ Moradi Marjaneh, Z.~G.~Rakhmatullina, M.~Ebrahimi-Loushab, D.~Saadatmand, V.~A.~Gani, P.~G.~Kevrekidis, and S.~V.~Dmitriev, {\it Collision of $\varphi^4$ kinks free of the Peierls-Nabarro barrier in the regime of strong discreteness}, \href{https://doi.org/10.1016/j.chaos.2020.109854}{{\it Chaos, Solitons and Fractals} {\bf 138} (2020) 109854}  [\href{https://arxiv.org/abs/1912.07953}{\tt arXiv:1912.07953}].

\bibitem{Manton.PRL.2021}
N.~S.~Manton, K.~Oleś, T.~Romańczukiewicz, and A. Wereszczyński, {\it Collective Coordinate Model of Kink-Antikink Collisions in $\phi^4$ Theory}, \href{https://doi.org/10.1103/PhysRevLett.127.071601}{{\it{ \ Phys.\ Rev.\ {\bf Lett}}}\ {\bf 127} (2021) 071601}  [\href{https://arxiv.org/abs/2106.05153}{\tt arXiv:2106.05153}].

\bibitem{Almeida.EPJC.2025}
F.~C.~E.~Lima, R.~Casana, and C.~A.~S.~Almeida, {\it Kinks and double-kinks in generalized $\phi^4$-and $\phi^4$-models}, \href{https://doi.org/10.1140/epjc/s10052-024-13651-3}{{{\it Eur.\ Phys.\ J.\ {\bf C}}} {\bf 84} (2025) 1266 }
[\href{https://arxiv.org/abs/2408.04761}{\tt arXiv:2408.04761}].

\bibitem{Moradi.JHEP.2017}
A.~Moradi~Marjaneh, V.~A.~Gani, D.~Saadatmand, S.~V.~Dmitriev, and K.~Javidan,
{\it Multi-kink collisions in the $\phi^6$ model},
\href{https://doi.org/10.1007/JHEP07(2017)028}{{JHEP} 07 (2017) 028}
[\href{https://arxiv.org/abs/1704.08353}{\tt arXiv:1704.08353}].

\bibitem{Gani.JOP.2020}
V.~A.~Gani, and A.~Moradi~Marjaneh, {\it Asymmetric kink solutions of hyperbolically deformed model}, \href{https://doi.org/10.1088/1742-6596/1690/1/012096} {{\it Journal of Physics: Conference Series} {\bf 1690} (2020) 012096}.

\bibitem{Saadatmand.EPJB.2022}
D.~Saadatmand, and A.~Moradi~Marjaneh, {\it Scattering of the asymmetric $\varphi^6$ kinks from a $\mathcal{PT}$-symmetric perturbation: creating multiple kink–antikink pairs from phonons}, \href{https://doi.org/10.1140/epjb/s10051-022-00405-x} {{\it Eur.\ Phys.\ J.\ {\bf B}} {\bf 95} (2022) 144}
[\href{arxiv.org/abs/2201.03277}{\tt arXiv:2201.03277}].

\bibitem{Adam.PRD.2022}
C.~Adam, K.~Oles, T.~Romanczukiewicz, and A.~Wereszczynski,
{\it Spectral walls in antikink-kink scattering in the $\phi^6$
model}, \href{https://doi.org/10.1103/PhysRevD.106.105027}{{\it Phys.\ Rev.\ {\bf D}} {\bf 106} (2022) 105027} [\href{https://arxiv.org/abs/2209.11479} {\tt arxiv:2209.11479}].

\bibitem{Saadatmand.CSF.2024}
D.~Saadatmand, A.~Moradi~Marjaneh, A.~Askari, and H. Weigel, {\it Phonons scattering off discrete asymmetric solitons in the absence of a Peierls-Nabarro potential}, \href{https://doi.org/10.1016/j.chaos.2024.114550}{{\it Chaos, Solitons and Fractals} {\bf 180} (2024) 114550}  [\href{https://arxiv.org/abs/2308.02322 }{\tt arXiv:2308.02322}].

\bibitem{Khare.PRE.2014}
A.~Khare, I.~C.~Christov, and A.~Saxena,
{\it Successive phase transitions and kink solutions in $\phi^8$, $\phi^{10}$, and $\phi^{12}$ field theories},
\href{https://doi.org/10.1103/PhysRevE.90.023208}{{{\it Phys.\ Rev.\ {\bf E}}} {\bf 90}  (2014) 023208}
[\href{https://arxiv.org/abs/1402.6766}{\tt arXiv:1402.6766}].

\bibitem{Blinov.JOP.2020}
P.~A.~Blinov, V.~A.~Gani, and A.~Moradi~Marjaneh, {\it From thin to thick domain walls: An example of the $\varphi^8$ model}, \href{https://doi.org/10.1088/1742-6596/1690/1/012082} {{\it Journal of Physics: Conference Series} {\bf 1690} (2020) 012082}.

\bibitem{khare2021explicit}
A.~Khare, A.~Duzgun, and A.~Saxena,
{\it Explicit Kink Solutions in Several One-Parameter Family of Higher Order Field Theory Models}, \href{https://doi.org/10.1142/S0217979221503240}{{\it Int.\ J.\ Mod. Phys.\ {\bf B}} {\bf 35} (2021) 2150324} [\href{https://arxiv.org/abs/2103.05145}{\tt arXiv:2103.05145}]. 

\bibitem{BCM} D.~Bazeia, J.~G.~F.~Campos, A.~Mohammadi, {\it Kink-antikink collisions in the $\phi^8$ model: short-range to long-range journey,}
\href{https://doi.org/10.1007/JHEP05(2023)116}{{JHEP} 05 (2023) 116}
[\href{https://arxiv.org/abs/2303.12482}{\tt arXiv:2303.12482}]. 

\bibitem{BBG}
D.~Bazeia, E.~Belendryasova, and V.~A.~Gani, {\it Scattering of kinks of the sinh-deformed $\phi^4$ model,} \href{https://doi.org/10.1140/epjc/s10052-018-5815-z}
{{\it Eur.\ Phys.\ J.\ {\bf C}} {\bf 78} (2018) 340}
[\href{https://arxiv.org/abs/1710.04993}{\tt arXiv:1710.04993}].

\bibitem{Moradi.CSF.2022}
A.~Moradi~Marjaneh, F.~C.~Simas, and D.~Bazeia
{\it Collisions of kinks in deformed $\varphi^4$ and $\varphi^6$ models},
\href{https://doi.org/10.1016/j.chaos.2022.112723}{{{\it Chaos, Solitons and Fractals}} {\bf 164}   (2022) 112723 }[\href{https://arxiv.org/abs/2207.00835}{\tt arXiv:2207.00835}].

\bibitem{Moradi.AnlPhys.2024}
A.~Moradi~Marjaneh, F.~C.~Simas, and D.~Bazeia
{\it Scattering of kinks in scalar-field models with higher-order self-interactions},
\href{https://doi.org/10.1016/j.aop.2024.169777}{{{\it Ann.\ Phys.}} {\bf 470}   (2024) 169777 }[\href{https://arxiv.org/abs/2402.00270}{\tt arXiv:2402.00270}].

\bibitem{Campbell.dsG.1986}
D. K.~Campbell, M.~Peyrard, and P.~Sodano, {\it Kink-antikink interactions in the double sine-Gordon equation}, \href{https://doi.org/10.1016/0167-2789(86)90019-9} {{\it Physica} {\bf D 19} (1986) 165}.

 \bibitem{Peyravi.EPJB.2009}
M.~Peyravi, A.~Montakhab, N.~Riazi, and A.~Gharaati, {\it Interaction Properties of the Periodic and Step-like Solutions of the Double-Sine-Gordon Equation}, \href{https://doi.org/10.1140/epjb/e2009-00331-0} {{\it Eur.\ Phys.\ J.\ {\bf B}} {\bf 72} (2009) 269}
[\href{https://arxiv.org/abs/0802.2776}{\tt arXiv:0802.2776}]. 

 \bibitem{Gani.EPJC.2018}
V.A.~Gani, A.~Moradi~Marjaneh, A.~Askari, E.~Belendryasova, and D.~Saadatmand,
{\it Scattering of the double sine-Gordon kinks},
\href{https://doi.org/10.1140/epjc/s10052-018-5813-1}{{\it Eur.\ Phys.\ J.\ {\bf C}} {\bf 78} (2018) 345}
[\href{https://arxiv.org/abs/1711.01918}{\tt arXiv:1711.01918}].

 \bibitem{Zolotaryuk.EPJC.2018}
Y.~Zolotaryuk, IO.~Starodub,
{\it Moving Embedded Solitons in the Discrete Double Sine-Gordon Equation},
\href{https://doi.org/10.1007/978-3-319-72218-4_13}{{\it Nonlinear Systems, Vol.2, Understanding Complex Systems.} Springer, Cham (2018)}.

\bibitem{Belendryasova.JPCS.2019}
E.~Belendryasova, V.A.~Gani, A.~Moradi~Marjaneh, D.~Saadatmand, and A.~Askari,
{\it A new look at the double sine-Gordon kink-antikink scattering},
\href{https://doi.org/10.1088/1742-6596/1205/1/012007}{{\it J.\ Phys.: Conf.\ Ser.} {\bf 1205} (2019) 012007}
[\href{https://arxiv.org/abs/1810.00667}{\tt arXiv:1810.00667}].

\bibitem{Yerin.PRB.2021}
Yuriy~Yerin, and Stefan-Ludwig~Drechsler, {\it Phase solitons in a weakly coupled three-component superconductor}, \href{https://doi.org/10.1103/PhysRevB.104.014518}{{\it{ \ Phys.\ Rev.\ {\bf B}}}\ {\bf 104} (2021) 014518}  [\href{https://arxiv.org/abs/2103.17000}{\tt arXiv:2103.17000}].

\bibitem{Hadipour.PLA.2020}
F.~Hadipour, D.~Saadatmand, M.~Ashhadi, A.~Moradi~Marjaneh, I.~Evazzade, A.~Askari, and S.~V.~Dmitriev,
{\it Interaction of phonons with discrete breathers in one-dimensional chain with tunable type of anharmonicity},
\href{https://doi.org/10.1016/j.physleta.2019.126100}{{\it Phys.\ Lett.\ {\bf A}} {\bf 384} (2020) 126100}.

\bibitem{DB-2022}Fabiano C.~Simas, K.~Z.~Nobrega, and D.~Bazeia, {\it Bifurcation and chaos in one dimensional chains of small particles,} \href{https://doi.org/10.1016/j.chaos.2022.112387}{{\it Chaos, Solitons and Fractals} {\bf 161} (2022) 112387}  [\href{https://arxiv.org/abs/2206.12734}{\tt arXiv:2206.12734}].

\bibitem{Simas.JHEP.2020}
F.~C.~Simas, F.~C.~Lima, K.~Z.~Nobrega, and A.~R.~Gomes,
{\it  Solitary oscillations and multiple antikink-kink pairs in the double sine-Gordon model},
\href{https://doi.org/10.1007/JHEP12(2020)143 }{{JHEP} 12 (2020) 143}
[\href{https://arxiv.org/abs/2007.12318}{\tt arXiv:2007.12318}].

 \bibitem{Bazeia.EPJC.2017}
D.~Bazeia, and D.~C.~Moreira,
{\it From sine-Gordon to vacuumless systems in flat and curved spacetimes},
\href{https://doi.org/10.1140/epjc/s10052-017-5458-5}{{\it Eur.\ Phys.\ J.\ {\bf C}} {\bf 77} (2017) 884}
[\href{https://arxiv.org/abs/1703.06363}{\tt arXiv:1703.06363}].

\bibitem{Speight.Nonlinearity.1997}
J.M.~Speight,
{\it A discrete $\phi^4$ system without a Peierls--Nabarro barrier},
\href{https://doi.org/10.1088/0951-7715/10/6/010}{Nonlinearity {\bf 10}, 1615 (1997)}
[\href{https://arxiv.org/abs/patt-sol/9703005}{\color{blue}\tt arXiv:patt-sol/9703005}].

\bibitem{kevrekidis.PhysicaD.2003}
P.G.~Kevrekidis,
{\it On a class of discretizations of Hamiltonian nonlinear partial differential equations},
\href{https://doi.org/10.1016/S0167-2789(03)00153-2}{Physica D {\bf 183}, 68 (2003)}.

\bibitem{Flach.PhysRep.2008}
S.~Flach, A.V.~Gorbach,
{\it Discrete breathers --- Advances in theory and applications},
\href{https://doi.org/10.1016/j.physrep.2008.05.002}{Phys.\ Rep.\ {\bf 467}, 1 (2008)}.

\bibitem{Zhanna.IOPconf.2018}
Zh.~G.~Rakhmatullina, P.~G.~Kevrekidis and S.~V.~Dmitriev, {\it Non-symmetric kinks in Klein-Gordon chains free of the Peierls-Nabarro potential,} \href{https://doi.org/10.1088/1757-899X/447/1/012057} {{\it  IOP Conf. Series: Materials Science and Engineering.} {\bf 447} (2018) 012057}

\bibitem{dmitriev.JPA.2005}
S.V.~Dmitriev, P.G.~Kevrekidis, N.~Yoshikawa,
{\it Discrete Klein--Gordon models with static kinks free of the Peierls--Nabarro potential},
\href{https://doi.org/10.1088/0305-4470/38/35/002}{J.\ Phys.\ A: Math.\ Gen.\ {\bf 38}, 7617 (2005)}
[\href{https://arxiv.org/abs/nlin/0506001}{\color{blue}\tt arXiv:nlin/0506001}].

\bibitem{BLM}
D.~Bazeia, L.~Losano, and J.M.C.~Malbouisson, {\it  Deformed defects}, \href{https://doi.org/10.1103/PhysRevD.66.101701}{Phys.\ Rev.\ D {\bf 66} (2002) 101701(R)} [\href{https://arxiv.org/abs/hep-th/0209027}{\color{blue}\tt arXiv:hep-th/0209027}]. 

\bibitem{DA}D.~Bazeia and L.~Losano, {\it Deformed defects with applications to braneworlds}, \href{https://doi.org/10.1103/PhysRevD.73.025016}{Phys.\ Rev.\ D {\bf 73} (2006) 025016} [\href{https://arxiv.org/abs/hep-th/0511193}{\color{blue}\tt arXiv:hep-th/0511193}]. 

\bibitem{DB} A. Khare and A. Saxena, {\it Novel Deformation Function Creating or Destroying any Number of Even Kink Solutions,} \href{https://doi.org/10.1016/j.physleta.2021.127830}{{Phys.\ Lett.\ {\bf A}} {\bf 424} (2022) 127830} [\href{https://arxiv.org/abs/2107.07048}{\color{blue}\tt arXiv:2107.07048}].

\bibitem{DC}P.~A.~Blinov, T.~V.~Gani, V.~A.~Gani, {\it Deformations of Kink Tails,} \href{https://doi.org/10.1016/j.aop.2021.168739}{{{ Ann.\ Phys.}} {\bf 437}   (2022) 168739 }[\href{https://arxiv.org/abs/2008.13159}{\tt arXiv:2008.13159}].

\bibitem{vasha}O. Albayrak and T. Vachaspati, {\it {Creating kinks with quantum mediation}}, Phys. Rev. D 109 (2024) 036001. [arXiv:2308.01962]
\bibitem{MM01}J.~Sampaio, J.~Grollier, P.~J.~Metaxas, {\it Domain Wall Motion in Nanostructures.} Chapter 8. In  {\it Magnetism of Surfaces, Interfaces, and Nanoscale Materials.} Edited by R. E. Camley, Z. Celinski,
R. L. Stamps. Handbook of Surface Science, 5 (2015) 335.

\bibitem{MM02}
M.~D.~DeJong and K.~L.~Livesey, {\it Analytic theory for the switch from Bloch to Néel domain wall in nanowires with perpendicular anisotropy,}  \href{https://doi.org/}{{Phys.\ Rev.\ B {\bf 92} (2015) 214420}} [\href{https://arxiv.org/abs/1510.07673}{\color{blue}\tt arXiv:1510.07673}].  

\end{thebibliography}
\end{document}